\documentclass[11pt,a4paper]{article}
\pdfoutput=1
\usepackage{jheppub}
\usepackage{hyperref}
\usepackage{setspace}
\usepackage{amsfonts,amssymb,epsfig,verbatim,mathbbol,amstext,amsmath,psfrag}
\usepackage{graphicx}
\usepackage[latin2]{inputenc}
\usepackage{t1enc}
\usepackage{amsmath}
\usepackage{subfigure}
\usepackage{dsfont}
\usepackage{slashed}
\usepackage{multirow}
\usepackage{lscape}
\usepackage{wrapfig}
\usepackage{array}
\usepackage{mciteplus}
\usepackage{enumerate}

\usepackage{colordvi}
\usepackage{lscape}
\usepackage{rotfloat}
\usepackage{rotating}
\usepackage{color}
\usepackage[bottom]{footmisc}
\usepackage{MnSymbol}

\usepackage{cmap}

\newcommand{\dd}{\textmd{d}}
\newcommand{\be}{\begin{equation}}
\newcommand{\ee}{\end{equation}}

\newcommand{\Z}{\mathcal{Z}}

\newcommand{\D}{\mathcal{D}}

\renewcommand{\O}{\mathcal{O}}
\newcommand{\expv}[1]{\left \langle #1 \right \rangle}

\newcommand{\tr}{\textmd{tr}\,}

\long\def\symbolfootnote[#1]#2{\begingroup%
\def\thefootnote{\fnsymbol{footnote}}\footnote[#1]{#2}\endgroup} 

\usepackage{tikz}
\usetikzlibrary{trees}
\usetikzlibrary{decorations.pathmorphing}
\usetikzlibrary{decorations.markings}
\usetikzlibrary{shapes.misc}
\usetikzlibrary{matrix}
\usetikzlibrary{decorations.pathreplacing}

\tikzset{
          photon/.style={decorate, decoration={snake,segment length=6pt, post length=0pt}, draw=black},
           nothing/.style={draw=white,very thick}
        }

\newcommand{\Budapest}{E\"otv\"os University, Theoretical Physics, P\'azm\'any P.\ s.\ 1/A, H-1117, Budapest, Hungary.}
\newcommand{\Regensburg}{Institute for Theoretical Physics, Universit\"at Regensburg, D-93040 Regensburg, Germany.}
\newcommand{\Frankfurt}{Institute for Theoretical Physics, Goethe Universit\"at Frankfurt, D-60438 Frankfurt am Main, Germany}
\newcommand{\Lendulet}{MTA-ELTE Lend\"ulet Lattice Gauge Theory Research Group. P\'azm\'any P.\ s.\ 1/A, H-1117, Budapest, Hungary.}
\newcommand{\Atomki}{Institute of Nuclear Research of the Hungarian Academy of
Sciences, Bem t\'er 18/c, H-4026 Debrecen,  Hungary.}
\newcommand{\Bonn}{HISKP(Theory), University of Bonn, Nussallee 14-16, D-53115 Bonn, Germany}

\title{Landau levels in QCD}

\author[a]{F.~Bruckmann,}
\author[b]{G.~Endr\H{o}di,}
\author[c,d]{M.~Giordano,}
\author[c,d]{S.~D.~Katz,}
\author[e]{T.~G.~Kov\'acs,}
\author[f]{F.~Pittler,}
\author[a]{J.~Wellnhofer}

\affiliation[a]{\Regensburg}
\affiliation[b]{\Frankfurt}
\affiliation[c]{\Budapest}
\affiliation[d]{\Lendulet}
\affiliation[e]{\Atomki}
\affiliation[f]{\Bonn}

\emailAdd{falk.bruckmann@ur.de}
\emailAdd{endrodi@th.physik.uni-frankfurt.de}
\emailAdd{giordano@bodri.elte.hu}
\emailAdd{katz@bodri.elte.hu}
\emailAdd{kgt@atomki.mta.hu}
\emailAdd{pittler@hiskp.uni-bonn.de}
\emailAdd{jacob.wellnhofer@ur.de}

\abstract{
We present first evidence for the Landau level structure of Dirac 
eigenmodes in full QCD for nonzero background magnetic fields, 
based on first principles lattice 
simulations using staggered quarks. Our approach involves the identification 
of the lowest Landau level modes in two dimensions, where topological arguments
ensure a clear separation of these modes from energetically higher states, and an expansion 
of the full four-dimensional modes in the basis of these two-dimensional states.
We evaluate various fermionic observables including the quark condensate 
and the spin polarization in this basis to find how much 
the lowest Landau level contributes to them. The results 
allow for a deeper insight into the dynamics of quarks and gluons in 
background magnetic fields and may be 
directly compared to low-energy models of QCD employing the lowest Landau level
approximation.
}

\begin{document}

\maketitle

\section{Introduction}

Background magnetic fields give rise to a wide range of exciting
phenomena with applications in solid state physics, cosmology, 
neutron star physics and heavy-ion phenomenology, see the recent reviews~\cite{Kharzeev:2013jha,Miransky:2015ava}. 
Our knowledge about 
these phenomena is guided by the quantum mechanics of charged particles 
exposed to background magnetic fields. The motion in this setup
is restricted to circular orbits (or spirals) with quantized radii. These so-called Landau levels (LL)
are responsible for various effects in solid state physics that involve
the electric conductivity or the magnetic moment of the material: 
the quantum Hall effect, the de Haas-van Alphen effect or the 
Shubnikov-de Haas effect (see, e.g., Ref.~\cite{abrikosov1988fundamentals}). 
The notable features of the Landau spectrum are the separation of the levels 
proportionally to the magnitude $B$ of the magnetic field, and the 
degeneracy of the levels, proportional to the magnetic flux $\Phi$ of the field 
through the area of the system. In particular, for strong fields the lowest
Landau level (LLL) plays the dominant role for macroscopic physics, since  
higher Landau levels (HLLs) are too energetic to be excited. 
An additional consequence of the LL-structure is the dimensional reduction 
of the theory for strong fields, where the motion is restricted to be parallel to the 
magnetic field.

If $B$ is sufficiently large, a weak interaction between the 
charged particles only perturbs the Landau levels, but 
leaves the overall hierarchy intact, so that the LLL dominance still holds.
In this paper our aim is to investigate whether the concept of Landau levels 
can also be transferred to strongly interacting quantum field theories and to 
what extent the LLL dominance persists in this case. 
In particular, we are interested in Quantum Chromodynamics (QCD),
which describes the strong (color)  
interaction between quarks and gluons.
While gluons are electrically neutral, quarks possess electric charge
and thus couple  
directly to the background magnetic field. 
It is worth emphasizing that the composite particles (e.g.\ charged
pions) of QCD  
have been observed to exhibit Landau
levels~\cite{Bali:2011qj,Chang:2015qxa,Bali:2015vua}.  
While this is expected for these weakly coupled 
particles, such a hierarchy has never been seen on the level of quarks, which 
interact strongly among each other.
The question of what role quark LLs could
play is especially interesting around and above the finite temperature crossover
to the quark-gluon plasma, because quark degrees of freedom become more important
here.
     
The most pronounced, magnetic field-induced effect in QCD is the 
enhancement of dynamical chiral symmetry breaking in the vacuum of the 
theory~\cite{Gusynin:1994re,Gusynin:1995nb}.
This, so-called magnetic catalysis is one of the most important features of 
the interaction between quarks, gluons and the magnetic field and has
a strong impact on the phase structure of QCD. It is widely believed
that the Landau level-structure of the theory 
-- in particular, the dimensional reduction for strong fields -- 
is responsible for magnetic catalysis. This expectation is backed up by 
calculations in various low-energy approximations, effective theories
and perturbative approaches  
to QCD. For recent reviews, we refer the reader to
Refs.~\cite{Shovkovy:2012zn,Andersen:2014xxa,Miransky:2015ava}. 
In addition, a convenient approximation exploiting the separation 
between the LLL and the HLLs
is to neglect 
all higher levels and only keep contributions from the LLL. This 
is the LLL approximation, which is widely employed, see, e.g.,
Refs.~\cite{Fukushima:2011nu,Leung:2005xz,Ferrer:2013noa,Fayazbakhsh:2010gc,Blaizot:2012sd,Ferrer:2014qka,Fayazbakhsh:2010bh,Fukushima:2015wck}. While 
the approximation may be justified for strong fields, neglecting the  
contributions from the HLLs results in systematic effects that are difficult to 
estimate~\cite{Kojo:2013uua,Mueller:2014tea,Braun:2014fua,Mueller:2015fka}. 
Notice that certain observables are special in this context as only
the LLL contributes to them: this is the case for anomalous
currents~\cite{Fukushima:2008xe} and for  
spin polarizations~\cite{Frasca:2011zn,Bali:2012jv} (see below).

The Landau level structure has further striking consequences: 
for vector mesons, the LLL carries a negative contribution to the energy that
has been speculated to turn the charged $\rho$ meson massless and,
accordingly, the QCD vacuum into  
a superconductor~\cite{Chernodub:2010qx}.
At high baryonic density and low temperature the gradual enhancement
of the Fermi energy results  
in a consecutive filling of the individual Landau levels and related 
oscillations.
The characteristic filling of the LLL was found 
to remain stable 
against color interactions using holography~\cite{Preis:2010cq}.

Yet another motivation to understand the role of Landau levels comes from the structure of 
the QCD phase diagram for nonzero magnetic fields. 
Lattice simulations have revealed~\cite{Bali:2011qj,Bali:2012zg,Endrodi:2015oba} (see also Ref.~\cite{Bornyakov:2013eya}) that around the deconfinement/chiral 
symmetry restoration transition of QCD,
the quark condensate is reduced by the magnetic field (inverse magnetic catalysis) -- 
an unexpected result if we compare it to the discussion above about the robust nature of magnetic 
catalysis. The impact of the LLL for inverse magnetic catalysis has been addressed, e.g., in 
Ref.~\cite{Ferrer:2014qka}. 
For a review on approaches to describe this phenomenon, see 
Refs.~\cite{Fraga:2012rr,Andersen:2014xxa}. 

In this paper we identify, for the first time, the Landau
level-structure of the quark Dirac operator on the lattice. After
defining the Landau levels in detail in Sec.~\ref{sec:LL}, we describe
our method to separate the lowest Landau level and the higher Landau
levels in two and in four dimensions. In Sec.~\ref{sec:observables} we
define the LLL-contribution to certain QCD observables including the
quark condensate and the spin polarization. This is followed by
Sec.~\ref{sec:results}, where we quantify the difference between the
LLL and the full theory for various magnetic fields and
temperatures. The observables and their divergences are calculated analytically in the  
free case in the appendices. Finally, Sec.~\ref{sec:summary} contains 
our conclusions.
Our preliminary results have been published in Ref.~\cite{Bruckmann:2016bcl}.

\section{Landau levels}
\label{sec:LL}

First of all we need to define Landau levels more specifically.
It is instructive to begin the discussion in two spatial dimensions and then 
proceed to the physical case of $3+1$ space-time dimensions. In addition, 
for each dimensionality we first describe the levels in the free theory, where 
quarks only interact with the magnetic field but not with gluons. Then, by 
switching on the strong interactions we can analyze whether the 
levels remain intact or if they are mixed.

\subsection{Two dimensions}

Let us consider a quark with electric charge $q$
that interacts with a background magnetic field 
$B$ but is otherwise free. 
In the following we will refer to this simply as the ``free case''. 
We work with natural units $c=\hbar=k_B=1$
and assume for simplicity $q>0$, $B>0$ and that the magnetic field
points in the $z$ direction.
In a finite periodic box of area $L^2$ in the $x-y$ plane, the flux of the magnetic field 
is quantized~\cite{'tHooft:1979uj,AlHashimi:2008hr} so that for the
flux quantum $N_b$ the following condition is satisfied: 
\be
N_b \equiv \frac{qB\,L^2}{2\pi} \in \mathbb{Z}\,.
\label{eq:quant}
\ee
The two-dimensional Dirac equation for such a background involves a coupling 
of $B$ both to the spin $\sigma_z$ and to the angular momentum
$L_z$ of the quark. These operators have quantized eigenvalues
$s_z=\pm1/2$ and $L_z=(2l+1)$ with $l\in \mathbb{Z}^+_0$. 
The eigenvalues of the massless Dirac operator (times $i$) will be referred to as energy levels.
The squared energies $\lambda_n^2$
and their degeneracy $\nu_n$ read  
\be
\lambda^2_n = qB \cdot ( 2l+1-2s_z)
= qB \cdot 2n
,\quad\quad 
\nu_n = N_b \cdot N_c \cdot (2-\delta_{n,0})\,,
\label{eq:freecase}
\ee
where we combined the angular momentum and spin into a single quantum
number $n\in\mathbb{Z}^+_0$ and $N_c=3$ denotes the number of colors.  
These levels are called Landau levels and $n$ is the Landau index.
Notice that since the contribution of the lowest angular momentum is
exactly canceled by $s_z=1/2$, the energy of the lowest Landau level
(LLL) with $n=0$ is zero independently of $B$. In addition, the LLL is 
the only level that has well-defined spin -- 
for our positively charged quark the spin is aligned with the magnetic field, 
$s_z=1/2$ (and the angular momentum $l$ vanishes). 
In contrast, higher Landau levels (HLLs) have no definite spin.
In the following we will index the eigenmodes either by the pair $(n,\alpha)$ 
with $n$ labelling the Landau levels and $0< \alpha\le \nu_n$ 
labelling the degenerate modes within each level, or simply by an integer $i$ running over 
all the modes (ordered according to the eigenvalues).

\begin{figure}[t]
 \centering
 \includegraphics[width=8cm]{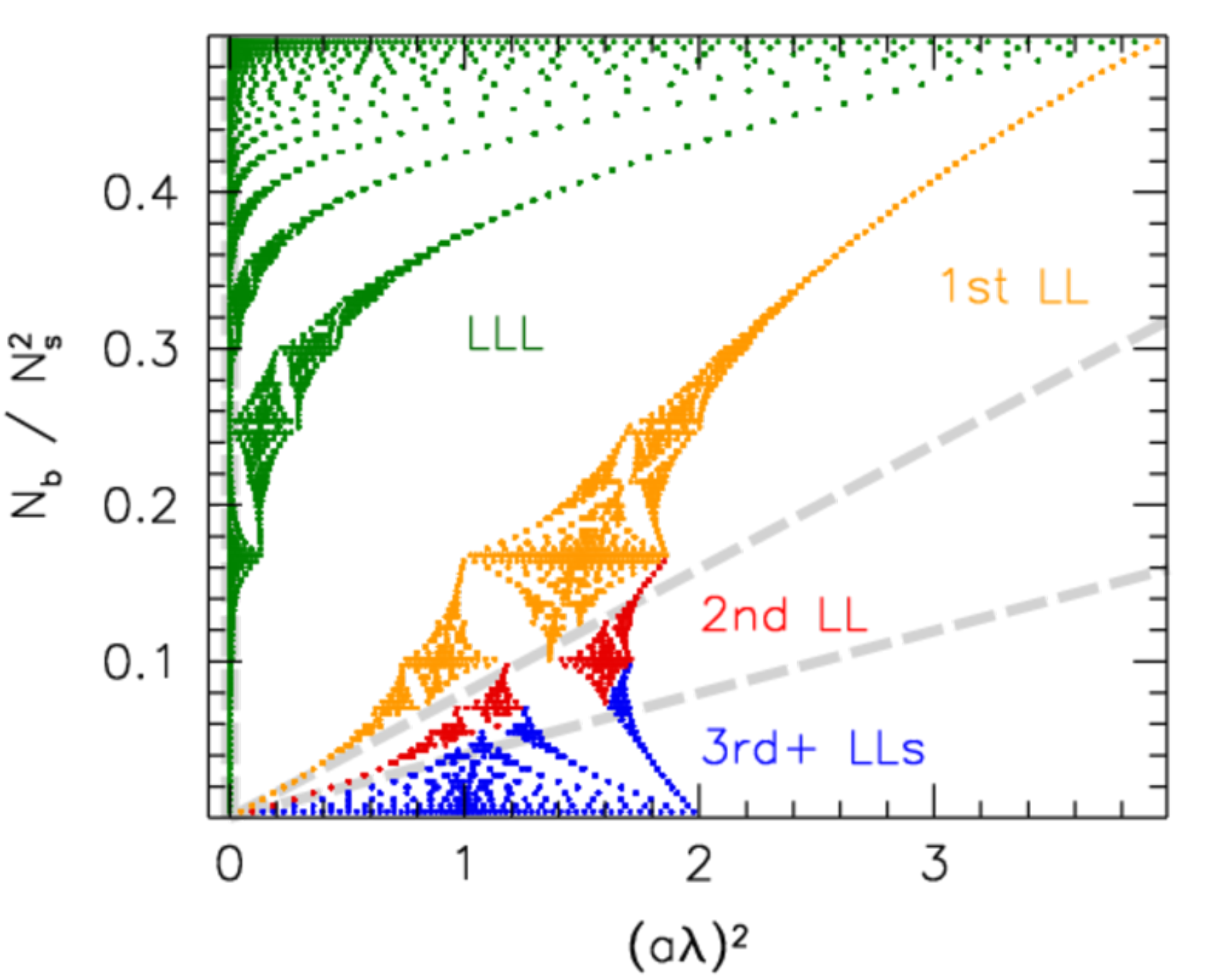}\quad\quad
 \includegraphics[width=8cm]{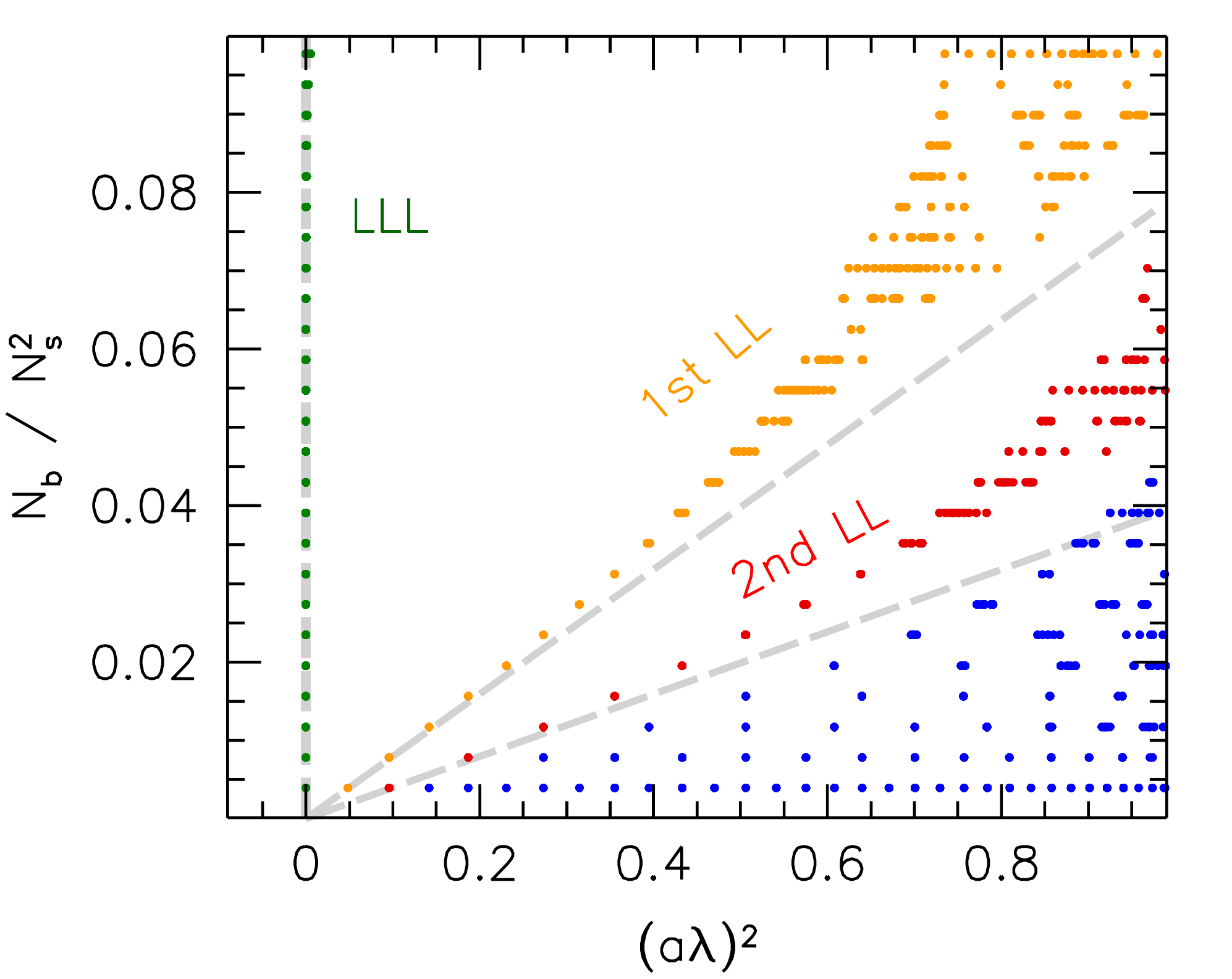}
 \caption{\label{fig:gap}Classification of the lattice eigenvalues according to continuum 
Landau level degeneracies. The left panel shows the complete spectra of the free two-dimensional Dirac operator, 
while in the right panel we zoom into the region around the origin, where the 
continuum Landau levels (gray dashed lines) are approached. 
Note that the latter show up as linear curves because the 
horizontal axis is $(a\lambda)^2$. 
}
\end{figure}

Next, we discretize space on a symmetric lattice with $N_s^2$ points
and a lattice spacing $a$ using the staggered Dirac operator. 
This formulation entails a twofold doubling of the squared eigenvalues. 
In addition, the lattice puts an upper limit $qB_{\rm max}=2\pi/a^2$ on the
allowed maximal magnetic field and 
the quantization condition~(\ref{eq:quant}) becomes
\be
N_b= \frac{qB \,(aN_s)^2}{2\pi}=0,1,\ldots,N_s^2.
\ee
The spectrum in this setting is shown in the left panel of Fig.~\ref{fig:gap}. 
The discretized system is near the continuum limit if the lattice is 
sufficiently fine to resolve the magnetic field: $a^2qB\ll 1$, i.e.\ 
$N_b/N_s^2\ll 1$. 
The right panel of Fig.~\ref{fig:gap} shows that this is indeed the case: for low flux quanta 
the eigenvalues of the lattice Dirac operator are on top of the continuum 
curves~(\ref{eq:freecase}). For higher values of $N_b$, the Landau level
hierarchy is broken by discretization artefacts so that the spectrum spreads 
around the continuum energies. This spread proceeds in an apparently recursive manner, with
the large-scale structure of the spectrum being repeated on ever smaller scales. The so emerging 
fractal is a well-known object in solid state physics and is called Hofstadter's butterfly~\cite{Hofstadter:1976zz}.

The butterfly has many spectacular features, some of which also
persist (at least partially) if 
QCD interactions are switched on~\cite{Endrodi:2014vza}. 
Here we concentrate on one of these characteristics: the structure of 
the gaps in the spectrum. The color coding of the eigenvalues in the
left panel of Fig.~\ref{fig:gap} corresponds to 
the continuum degeneracy~(\ref{eq:freecase}) -- ordering the eigenvalues 
according to their magnitude, the first $\nu_0\times 2 = N_cN_b\times 2$ entries are
assigned to the lowest (zeroth) LL, the next $\nu_1\times 2 = 2N_cN_b\times 2$ entries
to the first LL and so on. 
The factor of two is included to take into account the twofold fermion doubling 
mentioned above. 
Interestingly, this classification exactly coincides with the
separation in terms of the gaps. 
Another feature of the lattice spectrum is that the eigenvalues are
always below their corresponding continuum Landau levels -- with the
exception of the zeroth level, see the left panel of Fig.~\ref{fig:gap}.

\begin{figure}[t]
 \centering
 \mbox{
 \includegraphics[width=8cm]{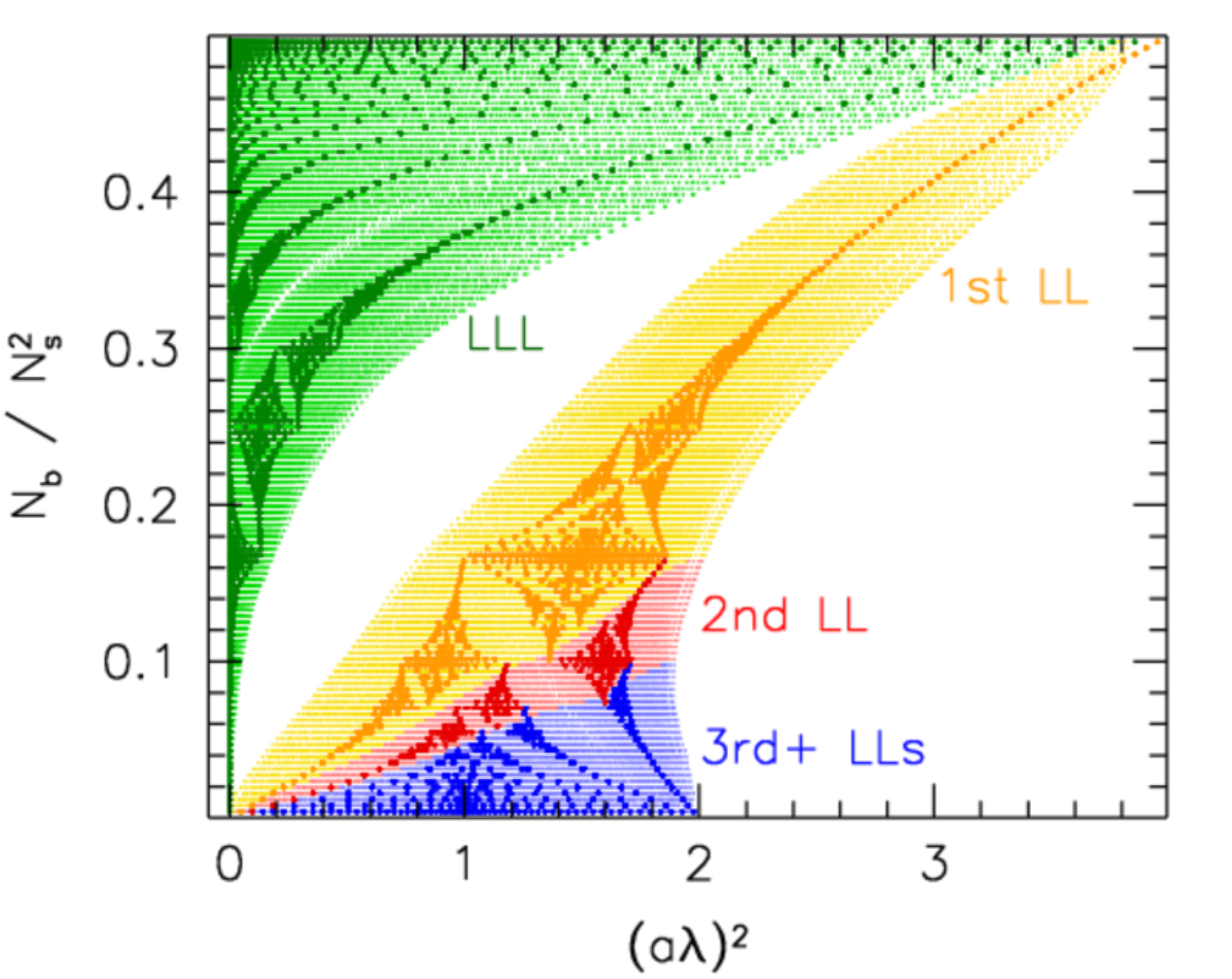}
 \includegraphics[width=9cm]{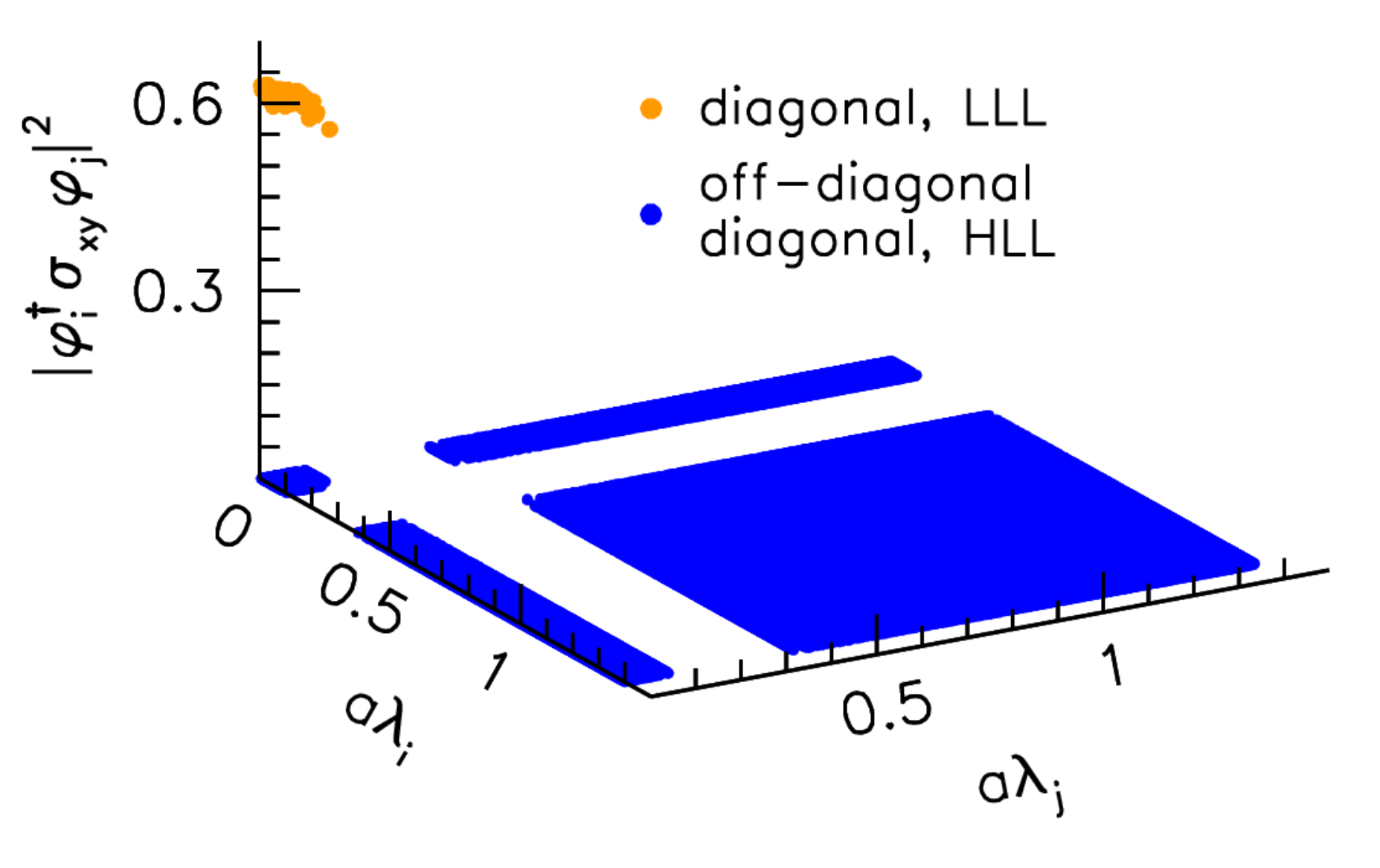} }
 \caption{\label{fig:spinxy}
 Left panel: the spectrum of the two-dimensional Dirac operator in the interacting case 
 -- evaluated on one slice of a typical four-dimensional 
gauge configuration (for details see the text). For comparison, the free-case eigenvalues from \protect Fig.~\ref{fig:gap} 
are also included. 
 Right panel: the absolute value square of the matrix elements of the relativistic spin operator
   $\sigma_{xy}$ in the basis of the  
 two-dimensional eigenmodes. Notice the separation of the LLL modes 
 from the HLL states by the gap (white region in the bottom plane) and the 
 very different matrix elements of $\sigma_{xy}$ on the two set of modes.  
 }
\end{figure}

Next we switch on QCD interactions by taking one $x-y$ slice of a
four-dimensional QCD gauge configuration
and inserting the links in the two-dimensional 
staggered Dirac operator $\slashed{D}_{xy}$. 
Thereby two new scales are introduced in the system: the strong scale 
$\Lambda_{\rm QCD}$ and the temperature $T$. In particular, here 
we consider a $16^3\times 4$ lattice from an ensemble generated 
at $T\approx 400 \textmd{ MeV}$. Notice that the minimal magnetic fields (i.e.\ small $N_b$) 
are then comparable to $\Lambda_{\rm QCD}^2$ and to $T^2$ so that a nontrivial competition
between these scales is expected to take place.
The so obtained spectrum is shown in the left panel of Fig.~\ref{fig:spinxy}, 
revealing that -- as expected -- the butterfly is smeared out by the
color interactions.  
Nevertheless, two crucial aspects of the lattice spectrum remain unaltered: 
a) the distinct \textit{presence of the largest gap} and b)
the correspondence of the left and right hand sides of the gap to LLL
and to HLLs, respectively, based on the \textit{continuum degeneracies}. 
These two features enable us to unambiguously separate the LLL from
HLLs in two-dimensional QCD.

Notice also that the smaller gaps between HLLs are closed by the
interactions, such that a similar distinction between, say, the first
and the second Landau level is not obvious. 
The LLL remains separate due to topological reasons. Namely, 
the topological charge in two dimensions is just the magnetic flux 
(even in the presence of non-Abelian interactions)
\be
Q^{\rm 2D}_{\rm top} = \frac{1}{2\pi} \int \dd^2 x\, F_{xy} = \frac{1}{2\pi}\,
L^2 \cdot qB = N_b\,, 
\label{eq:2dindexth}
\ee
and the usual four-dimensional notion of handedness is replaced by the spin 
direction, thus the index theorem entails that $Q^{\rm 2D}_{\rm top} =
N_\uparrow-N_\downarrow$ equals the difference of the number of
spin-up and spin-down polarized zero modes. In addition, in two
dimensions
the `vanishing theorem'~\cite{Kiskis:1977vh,Nielsen:1977aw,Ansourian:1977qe} 
ensures that either $N_\uparrow$ or $N_\downarrow$ is zero. Thus, for
$qB>0$ 
the
only states in the spectrum with definite spin have spin up
and according
to Eq.~(\ref{eq:2dindexth}) $N_b=N_\uparrow$. Indeed,
the LLL eigenvalues vanish in the continuum\footnote{In the staggered
  discretization the LLL modes are not real zero modes but are 
well separated from the HLL eigenvalues. In the overlap
formulation~\cite{Neuberger:1997fp,Neuberger:1998wv} these modes
become exact zero modes.},  
and their degeneracy is $N_b$ (for each color). 

To demonstrate that even in the presence of color interactions 
the LLL only accommodates spin-up states, in the right panel of 
Fig.~\ref{fig:spinxy} we plot the squared matrix elements  
$|\varphi_i^\dagger \sigma_{xy} \varphi_j|^2$ of the spin
operator\footnote{The staggered discretization of the spin operator is
  detailed in Ref.~\cite{Bali:2012jv}.}  
$\sigma_{xy} = \sigma_{z}$ for the down quark at a magnetic flux quantum 
$N_b=10$.
Besides the separation of the LLL modes ($i\le N_bN_c\times 2$) from the 
 HLL modes ($i > N_bN_c\times 2$), the two sets are also clearly distinguished by their 
 spin matrix element. In particular, 
we find that $\sigma_{xy}$ is almost perfectly diagonal in the eigenmode basis -- 
the off-diagonal matrix elements are below $10^{-4}$. For the diagonal 
elements, the HLL entries are also suppressed (below $10^{-2}$), while 
the LLL entries are much larger, in this case around $0.6$. In fact, 
the spin of the LLL modes approaches unity in the continuum limit. 
Thus, the classification of the
two-dimensional modes based on  
their mode number (LLL degeneracy) coincides with the
classification based 
on their spin.

\begin{figure}[t]
 \centering
 \includegraphics[width=7.5cm]{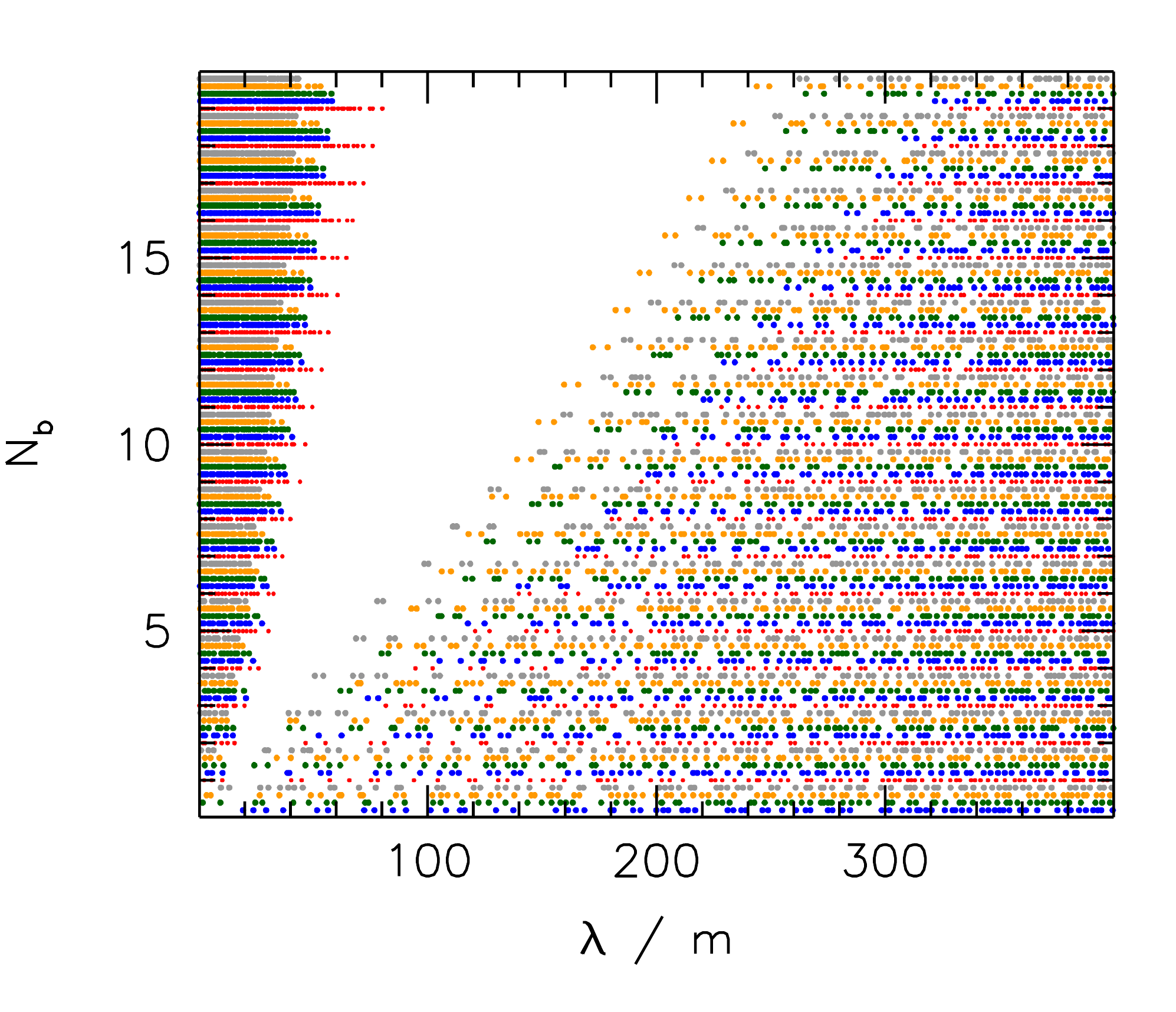} \quad
 \includegraphics[width=7.5cm]{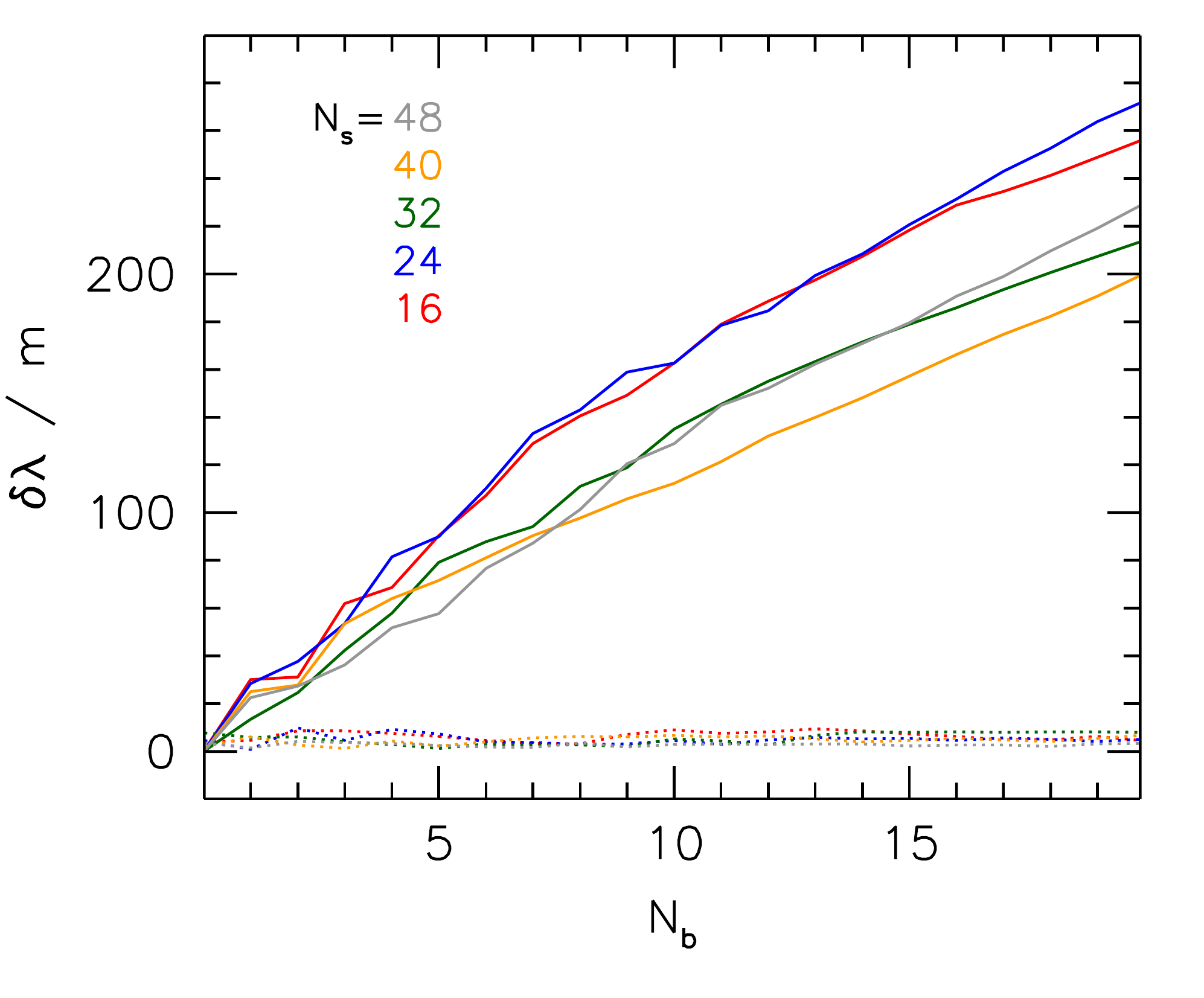}
 \caption{\label{fig:contgap} Left panel: the gap between the LLL and the 
 HLLs in physical units for 
 various lattice spacings in full two-dimensional QCD (the 
 different eigenvalue sets have been shifted vertically for better visibility).
 Right panel: the width of the gap (solid lines) 
 compared to the typical eigenvalue spacing
 just above the gap (dotted lines). The color coding of the left panel 
 matches that of the right panel.
 }
\end{figure}

It is therefore the index theorem that protects the LLL states from
mixing with HLL modes, resulting in the  
persistence of the gap even in the presence of QCD interactions.
To show that the above characteristics remain to hold in the continuum limit,
we plot the gap for
various lattice spacings in a fixed physical volume $L^2$
in the left panel of Fig.~\ref{fig:contgap}. 
The employed QCD configurations are two-dimensional slices of 
typical high-temperature ($T\approx400\textmd{ MeV}$) 
four-dimensional gauge configurations with aspect ratio
$N_s/N_t=4$ and $N_s=16\ldots 48$. The gap is shown this time
in physical units: the magnetic flux $N_b=qB L^2 /(2\pi)$ 
on the horizontal and the eigenvalue in units of the bare quark mass on the
vertical axis.\footnote{This choice of normalization is dictated by
  the fact that expressing the eigenvalues in units of the bare quark
  mass leads to a renormalization-group-invariant spectral
  density~\cite{Giusti:2008vb}, 
  and is thus required to obtain a meaningful continuum limit.}
Apparently, the gap edges remain  
well-defined also in the limit $a\to0$ (i.e.\ $N_t\to\infty$). To be more specific, in the
right panel of the same  
figure we plot the width $\delta\lambda$ 
of the gap as a function of $N_b$, together with the eigenvalue spacing 
just above the gap. We see that the gap width always largely
exceeds the typical spacing -- in other words, the gap at small flux 
quanta is indeed a well-defined physical structure that survives
the continuum limit. Notice moreover that as the continuum limit is approached,
the LLL states -- while having a fixed multiplicity $N_cN_b$ -- are
compressed towards zero, in accordance with their would-be-zero-mode nature.
Finally we remark that above we presented the pronounced features of the spectrum 
using high-temperature QCD ensembles, 
but our main conclusions remain unchanged if we use instead gauge configurations in the confined phase.

\subsection{Four dimensions}
\label{sec:fourd}

Next, we generalize the concept of Landau levels to four dimensions in
Euclidean spacetime. 
In the absence of color interactions, the Dirac equation for the 
$z$ and $t$ coordinates decouples from the Landau problem in the $x-y$ plane and
has free wave solutions with momenta $p_z$ and $p_t$. 
Thus, the eigenmodes factorize as 
$\psi_{n \alpha  p_z p_t} = \varphi_{n \alpha} \otimes e^{i p_z z}
\otimes  e^{i p_t t}$,  where $n$ labels the LL and $\alpha$ the degenerate 
modes within each level. 
The squared eigenvalues and their degeneracies read
\be
\lambda^2_{n p_z p_t} 
= qB \cdot 2n  +p_z^2+p_t^2
,\quad\quad 
\nu_{n p_z p_t} = 2N_b \cdot N_c \cdot (2-\delta_{n,0})\,.
\label{eq:freecase4d}
\ee
(Here we assumed strictly zero temperature, i.e.\ an infinite size in
the temporal direction.) 
Therefore, each Landau level has become an infinite tower $\psi_{n \alpha p_z p_t}$ 
of states, involving all the allowed momenta in the $z$ and $t$ directions. 
As a consequence, it is not possible anymore
to separate the LLL from the HLLs just by looking at the eigenvalues
$\lambda_{n p_zp_t}$. 
Clearly, we need to extract the Landau index $n$ from the eigenmode,
or, in other words,  
work with a projector $P$ that projects onto the subspace spanned by the 
modes with the lowest Landau index $n=0$,
\be
\textmd{continuum, non-interacting:}\quad
P=\sum_{p_z, p_t} \sum_\alpha \psi_{0 \alpha p_z p_t} \psi_{0 \alpha p_z p_t}^\dagger =
\sum_{\alpha} \varphi_{0\alpha} \varphi_{0\alpha}^\dagger \otimes \mathbb{1}_z \otimes
\mathbb{1}_t\,. 
\label{eq:projdef1}
\ee

On the lattice, the eigenmodes still factorize as 
$\psi_{i p_z p_t}=\varphi_i \otimes  \frac{1}{\sqrt{N_s}} e^{i p_z z}
\otimes  \frac{1}{\sqrt{N_t}} e^{i p_t t}$
(here $i$ runs over all the two-dimensional modes, see our remark after Eq.~(\ref{eq:freecase})).
Instead of using the plane wave basis in the $z$ and $t$ directions, we can also 
span the same space by using the coordinate basis consisting of states localized at a single value of $z$ and of $t$,
\be
\psi_{izt}(x,y,z',t') = \varphi_i(x,y) \otimes \delta_{zz'} \otimes \delta_{tt'}\,.
\label{eq:psiiztdef}
\ee
After ordering the $\varphi_i$ according 
to their eigenvalues, 
the first $N_cN_b\times 2$ two-dimensional states correspond to the
LLL 
(see Fig.~\ref{fig:gap}).  
Therefore, a valid way to rewrite~(\ref{eq:projdef1}) is to only
include these modes, 
\be
\textmd{lattice, non-interacting:}\quad
P = \sum_{i\le N_cN_b} \,\sum_{\rm doublers} \,\sum_{z, t} \psi_{i
  z t} \psi_{i z t}^\dagger =  
\sum_{i\le N_cN_b}\,\sum_{\rm doublers}  \varphi_i \varphi_i^\dagger
\otimes \mathbb{1}_z \otimes \mathbb{1}_t\,, 
\label{eq:projdef2}
\ee
where the sum over doublers appears due to the twofold doubling of
staggered fermions in two dimensions. 

If QCD interactions are switched on, the components of the four-dimensional 
Dirac operator $\slashed{D}=\slashed{D}_{xy}+\slashed{D}_{zt}$ in general 
do not commute, i.e.\ the eigenmodes do not factorize as in the free
case above. Nevertheless, we may still employ the basis of
the eigenstates $\varphi_i^{(z,t)}$ of $\slashed{D}_{xy}^{(z,t)}$ for each $x-y$ plane of the
lattice, labelled by the coordinates $z,t$. The factorized modes $\psi_{izt}$ are built up from these, similarly as in Eq.~(\ref{eq:psiiztdef}),
\be
\psi_{izt}(x,y,z',t') = \varphi_i^{(z,t)}(x,y) \otimes \delta_{zz'} \otimes \delta_{tt'}\,.
\label{eq:psiizt}
\ee
Thus, the projection in this setting reads 
\be
\textmd{lattice, interacting:}\quad
P=
 \sum_{i\le N_cN_b}\,\sum_{\rm doublers} \sum_{z,t} \psi_{izt}
\psi_{izt}^\dagger\,. 
\label{eq:projdef3}
\ee
This is the projector we will use in full four-dimensional QCD to pick
out the states corresponding to the LLL. 
Later we will also use the same 
construction but composed of the eigenmodes $\widetilde{\varphi}^{(z,t)}_i$ of 
the two-dimensional Dirac operator at {\it vanishing} magnetic field. 
Similarly as above, this uses $\widetilde{\psi}_{i z
  t}(x,y,z',t')=\widetilde{\varphi}_i^{(z,t)}(x,y) \otimes  \delta_{zz'}
\otimes 
\delta_{tt'}$ and reads
\be
\widetilde{P} =  \sum_{i\le
  N_cN_b}\,\sum_{\rm doublers}\sum_{z,t}  \widetilde{\psi}_{izt} 
\widetilde{\psi}_{izt}^{\,\dagger}\,. 
\label{eq:projdef_extra}
\ee
Once again, $\widetilde{P}$ involves the same number $N_cN_b\times 2$ of modes as 
$P$ does, 
but the modes are eigenstates of the $B=0$ Dirac operator.

Our numerical results will show that the four-dimensional modes of $\slashed{D}$
never correspond purely to the LLL or to a HLL but instead -- owing to
the mixing between the various $x-y$ planes via gluon fields in the
$z$ and $t$ directions -- have overlap both with $P$ and with its complement $1-P$. 
Nevertheless, for typical low-lying four-dimensional modes, 
there is a distinct jump in the overlap with $\psi_{izt}$
between $i=N_cN_b\times 2$ and $i=N_cN_b\times 2+1$ i.e.\ just at the
border of the LLL.

\begin{figure}[t]
 \centering
 \includegraphics[width=8cm]{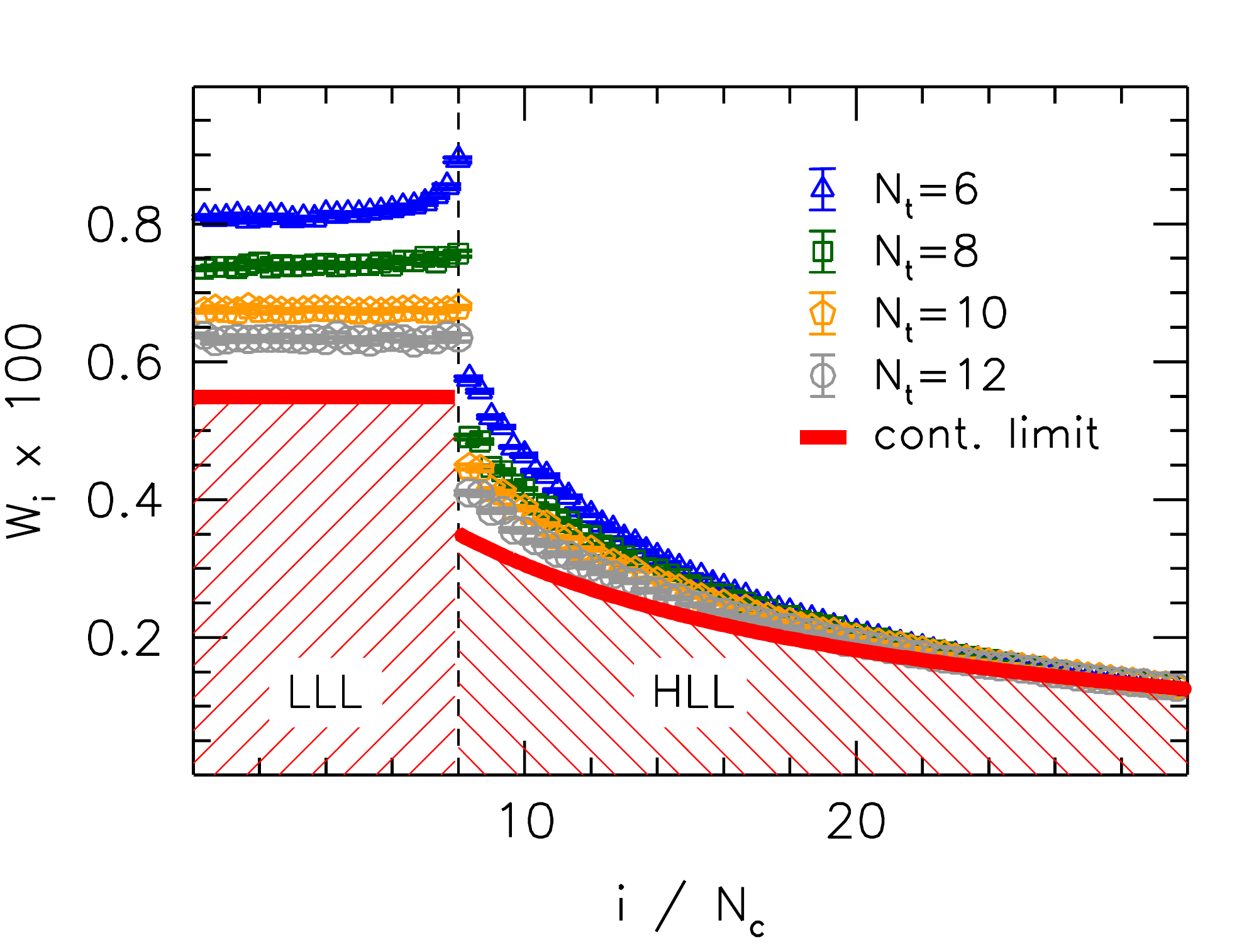}\quad
 \includegraphics[width=8cm]{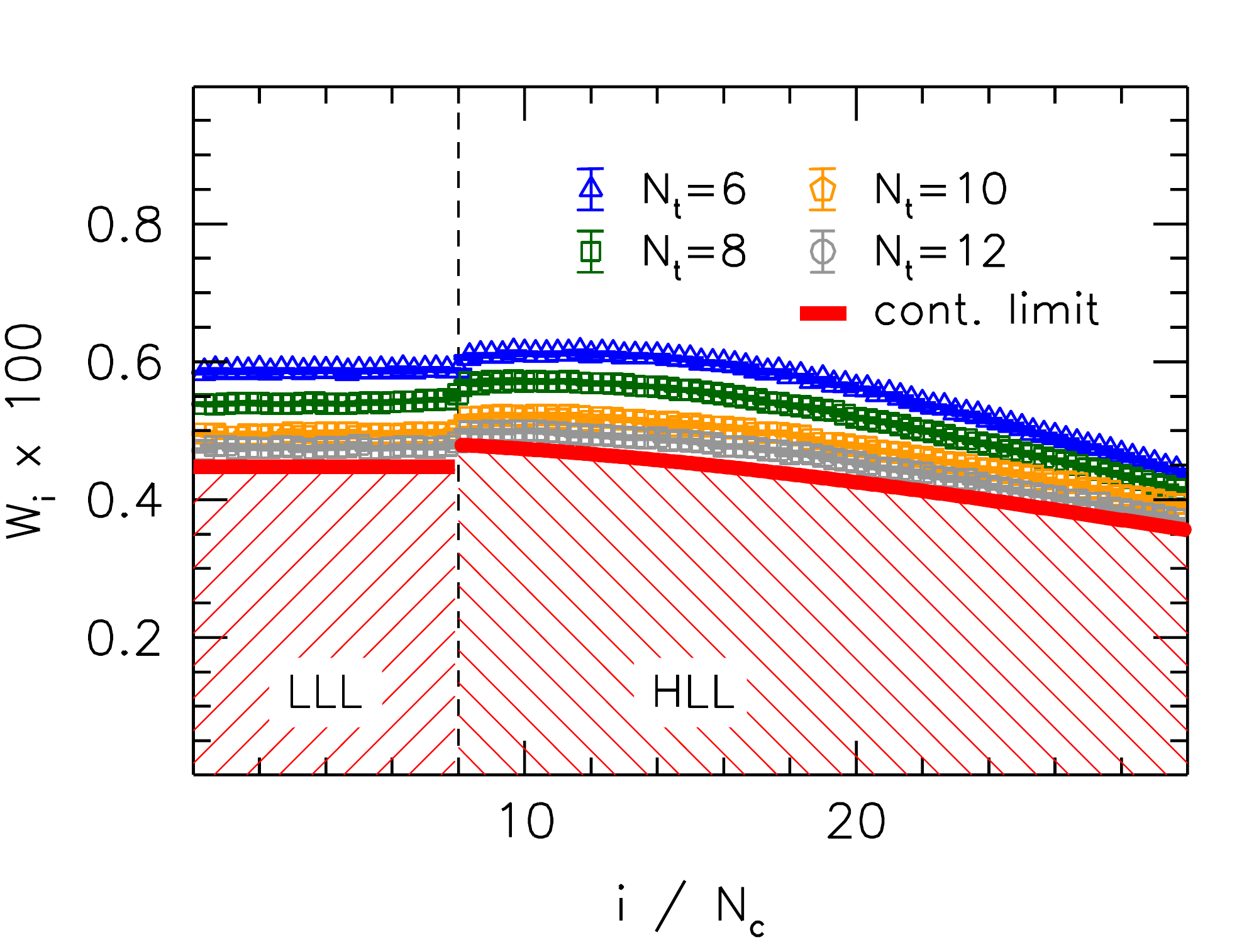}
 \caption{\label{fig:overlap}
 The overlap~(\protect\ref{eq:overlapdef}) of four-dimensional
 eigenmodes with the two-dimensional modes as a function of the index
 (in units of $N_c=3$) of the latter for a magnetic flux quantum $N_b=8$. 
 The left panel corresponds to low-lying four-dimensional 
 modes (with eigenvalue $220<\lambda/m<225$) while the right panel represents bulk modes
 ($535<\lambda/m<545$) on configurations generated at $T\approx400 \textmd{ MeV}$.
   }
\end{figure}

This is visualized in the left panel of Fig.~\ref{fig:overlap} for
normalized four-dimensional modes $\phi$ for the down quark ($q_d=-e/3$) 
at $T=400 \textmd{ MeV}$. We
define the overlap factor as 
\be
W_i(\phi) = \sum_{\rm doublers} \sum_{z,t} 
\big|\psi_{izt}^\dagger \phi\big|^2\,,
\label{eq:overlapdef}
\ee
where the sum also includes the two two-dimensional doublers 
and the scalar product $\psi_{izt}^\dagger \phi$ involves a sum over all 
lattice points. 
The completeness of the $\psi_{izt}$ modes ensures that 
the normalization is $\sum_i W_i(\phi)=\phi^\dag \phi = 1$. We average over four-dimensional modes in a
small spectral interval and over several gauge configurations. 
The magnetic flux used here is $N_b=8$, leading to a  
LLL degeneracy of $N_cN_b=24$. The left panel of Fig.~\ref{fig:overlap} reveals that 
low-lying modes $\phi$ tend to have 
larger overlap with two-dimensional LLL modes
than with HLL states. 
$W_i$ also remains constant in the LLL region $i\le N_bN_c$, suggesting the 
equivalence of all two-dimensional lowest Landau levels in this respect. 
This feature, together with the 
drastic downward jump at the end of the LLL region remains
pronounced 
even in the continuum limit.
 
In the right panel of Fig.~\ref{fig:overlap} we plot the same quantity, only 
this time $\phi$ are high-lying four dimensional modes that are expected to 
have less overlap with the LLL. Indeed, the pronounced downward jump becomes 
a slight upward jump, so that 
these modes can be rather 
thought of as being HLL-dominated.
We also mention that the structures visible in Fig.~\ref{fig:overlap} disappear 
for $B=0$ and the overlap becomes a smooth, monotonically decreasing function.

\begin{figure}[t]
 \centering
 \includegraphics[width=8cm]{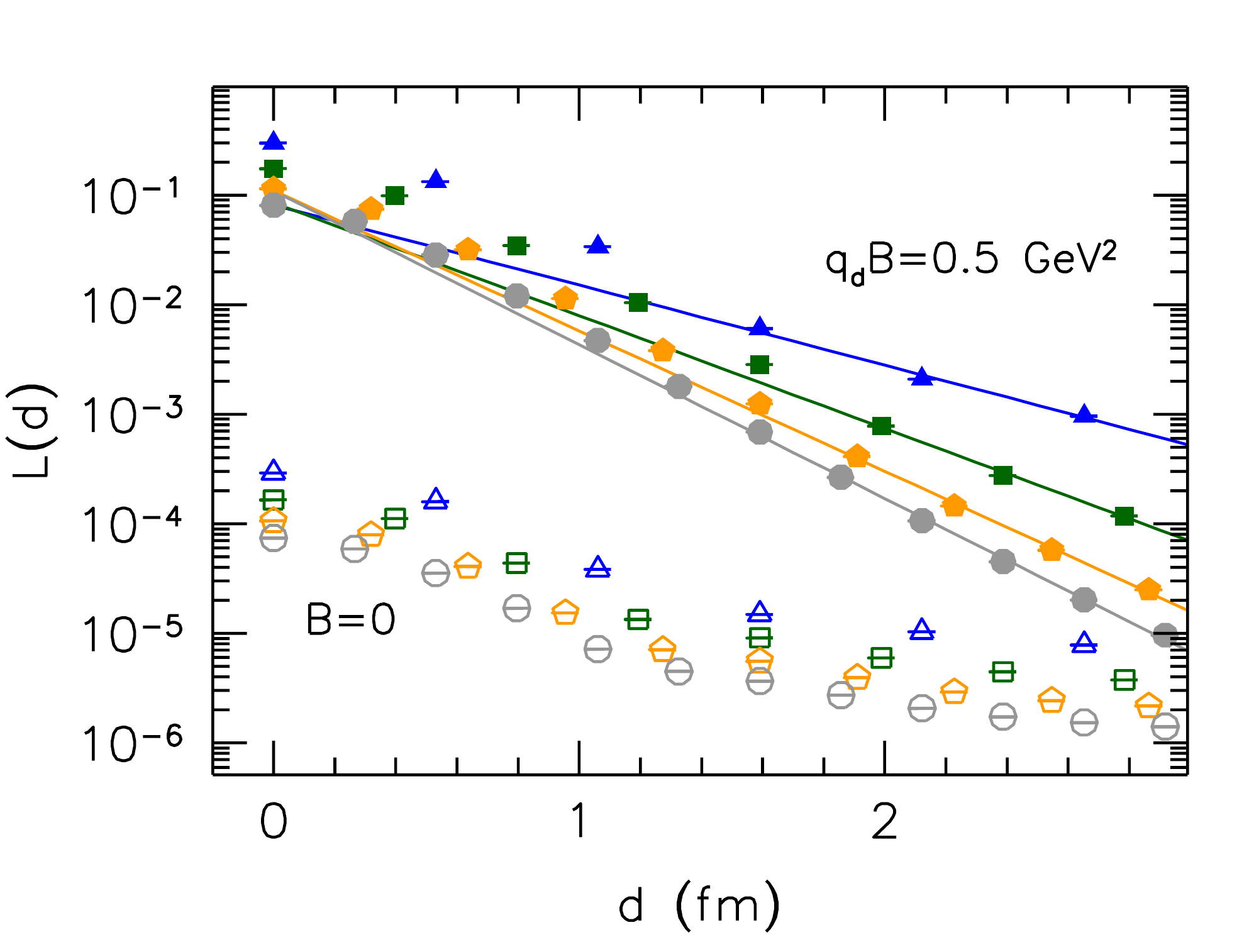}\quad
 \includegraphics[width=8cm]{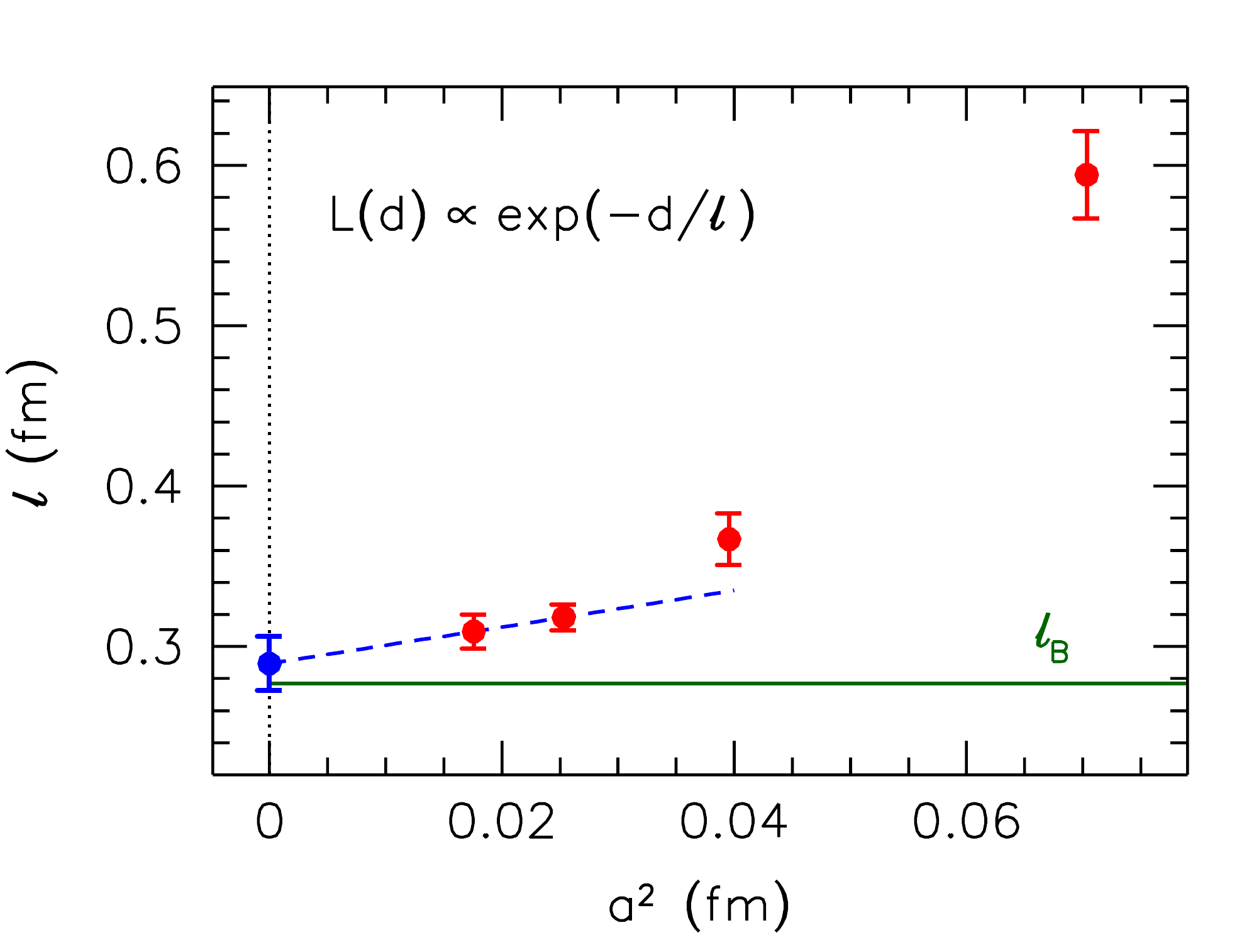}
 \caption{\label{fig:locality}Left panel: the expectation value $L(d)$ of the 
modulus of the vector obtained by projecting a localized source on the
LLL, as a function of the distance $d$ from the source, for a magnetic field 
$q_dB=0.5\textmd{ GeV}^2$. For the proper definition 
of $L(d)$ see the text. The different colors correspond to $N_t=6$, $8$, $10$ and $12$, 
from top to bottom. 
The exponential decay (solid lines) for $B>0$ demonstrates the locality
 (in the sense described in the text) 
 of the LLL-projected fermionic action. For $B=0$ (where we shifted the data vertically 
 for better visibility) no such decay is
 observed.
 Right panel: the continuum extrapolation of the decay length based on the 
 two finest lattices, compared to the expected 
 localization length $\ell_B=1/\sqrt{q_dB}$.} 
\end{figure}

Another important aspect regarding LLL-projected fermions is
the locality of the fermion action corresponding to the Dirac operator restricted 
to the LLL subspace. Note that already in the continuum, the lowest Landau level
spreads over a range $\ell_B=1/\sqrt{qB}$ in the plane perpendicular to the magnetic field
(see, e.g., Ref.~\cite{Miransky:2015ava}), 
so that the LLL-projected quark action involves (contrary to usual QCD) 
the product of quark fields smeared over the range $\ell_B$. 
A nontrivial check of our lattice construction is whether this localization range 
is reproduced. 
The original Dirac operator without the projection $P$ is ultralocal as it only uses nearest neighbor links.
Thus, for LLL-projected fermions we need to check the locality of the
projector itself. As can be seen directly from Eqs.~(\ref{eq:psiizt}) and~(\ref{eq:projdef3}),
$P$ is ultralocal in the $z$ and $t$
directions. To discuss
locality in the $x-y$ plane,  
we consider a source vector $\xi$ localized at the point $(x,y,z,t)$ and the vector 
$\psi=P\, \xi$ obtained by projecting with $P$. We measure 
$L(d)= \expv{\Vert
  \psi(x',y',z,t)  \Vert}$, with $d=\sqrt{(x-x')^2+(y-y')^2}$ the
distance from the source in the $x-y$ plane. 
As the left panel of Fig.~\ref{fig:locality}
reveals, this quantity 
falls off exponentially with the distance, signaling that interactions between 
sufficiently separated quark fields are indeed suppressed 
-- 
once the averaging over gluonic configurations is
performed. In the right panel of Fig.~\ref{fig:locality}
we perform the $a\to0$ extrapolation of the decay length and find that in the 
continuum limit it is indeed consistent with the expected value $\ell_B\approx 0.28 \textmd{ fm}$
for the magnetic field considered here. 
We emphasize that the locality of $P$ (in the sense described above) is a
highly nontrivial finding that arises from the interplay of the LLL
modes.\footnote{We mention that the locality of the action in {\it lattice units} 
(i.e., $\ell\to0 \textmd{ fm}$ for $a\to0$) for usual QCD is a prerequisite 
for the universality of the continuum limit. For the LLL-projected action 
the microscopic details of the discretization 
are damped by the magnetic localization length, present even in the continuum theory. 
Strictly speaking, the universality argument
can therefore only be applied for $B\to\infty$ 
where $\ell_B\to0$.} In general, the projection onto a subset of eigenmodes  
of the Dirac operator is a highly nonlocal object. We demonstrate this by
applying the same construction at  
$B=0$ -- building the projector $\widetilde{P}$ of
Eq.~(\ref{eq:projdef_extra}) from the lowest $N_bN_c\times2$
two-dimensional modes at {\it vanishing} magnetic field. The so  
obtained object does not appear to exhibit exponential decay, see
the left panel of Fig.~\ref{fig:locality}.  

\section{Observables}
\label{sec:observables}

Having prescribed the procedure to project an arbitrary four-dimensional mode 
to the LLL sector, we are in the position to test to what extent certain 
QCD observables are LLL dominated. 
We work with three quark flavors indexed by $f=u,d,s$.  
The observables can be derived from the QCD partition function $\Z$, which 
is written using the Euclidean path integral over gluon $A_\mu$ and 
quark fields $\Psi=(\psi_u,\psi_d,\psi_s)^{\top}$,
\be
\Z = \int \!\D A_\mu \,\D\bar\Psi \,\D\Psi \,e^{-S_G-S_F}, \quad\quad
S_F = \bar\Psi M \Psi = \sum_{f=u,d,s} \bar\psi_f M_f \psi_f\,,
\label{eq:Z}
\ee
where $S_G$ and $S_F$ are the gluonic and fermionic actions, respectively, 
and $M=\textmd{diag}(M_u,M_d,M_s)$ with flavor blocks 
$M_f=\slashed{D}_f+m_f$ denotes the quark matrix. 
The Dirac operator is flavor-dependent due to the different electric charges:
$q_u=-2q_d=-2q_s=2e/3$ with $e>0$ the elementary charge. 
Integrating out the fermion fields analytically, the expectation values of 
quark bilinears for the flavor $f$ read
\be
\expv{\bar\psi_f\Gamma\psi_f} = \frac{T}{4V} \frac{1}{\Z} \int \!\D
A_\mu \,e^{-S_G}\, {\det}^{1/4} [M]\; \tr[M_f^{-1}\Gamma\,] \,. 
\label{eq:exppsi}
\ee
Here, the rooting trick for staggered quarks 
is employed to reduce the number of flavors to three in the continuum limit 
and the division by the four-volume $V/T$ renders the observable intensive.
The details of our lattice setup, including the simulation algorithm, the
implementation of the magnetic field and the line of constant physics
to set the quark masses $m_u=m_d$ and $m_s$ are described in
Refs.~\cite{Borsanyi:2010cj,Bali:2011qj}. 

To find the LLL contribution to the observable~(\ref{eq:exppsi}), we
work with the projector of  
Eq.~(\ref{eq:projdef3}), which 
projects onto the subspace spanned by all of the two-dimensional LLL modes
(defined on two-dimensional slices corresponding to all values of $z$ and of $t$).
The projector is a block-diagonal matrix in flavor space,
$P=\textmd{diag}(P_u,P_d,P_s)$.  
The LLL projection amounts essentially to replacing the fermion matrix
$M$ by its projected version $PMP$. After integrating out the
fermions, $M$ shows up both in the trace and in the determinant in
Eq.~(\ref{eq:exppsi}). One can then consider the effect of the LLL
projection on the valence quarks in the operator in question -- represented by the trace -- as well as
on the sea quarks that characterize the distribution of the gluonic
configurations
-- represented by the determinant. In the present approach, we only insert the projector
in the valence sector. The sea contribution is considerably more
complicated to implement and we leave it to a forthcoming
study.\footnote{Here we mention only that if one wants to insert $P$
  in the determinant, one should also ``quench'' the HLL modes, i.e.,
  the correct replacement would be $M\to PMP + 1-P$.} 
Similarly, we only insert the magnetic field in the valence Dirac operator 
and exclude $B$ in the sea sector, i.e., 
for the generation of the gauge configurations. 
This implies that valence quarks feel the magnetic field and 
are projected to the LLL, 
while virtual sea quarks behave as if they were electrically neutral. 
We mention that the valence contribution is dominant for, e.g., the
quark condensate 
at low temperatures~\cite{DElia:2011koc} but not around the QCD transition~\cite{Bruckmann:2013oba}. This should be
kept in mind in  
the following.

Our definition of the full and the LLL projected quark bilinears (in the valence approximation) thus reads
\be
\begin{split}
\expv{\bar\psi_f\Gamma\psi_f}_B 
&= \frac{T}{4V}\frac{1}{\Z(0)} \int \!\D A_\mu \,e^{-S_g} \,{\det}^{1/4} [M(0)]\; \tr[M_f^{-1}(B)\Gamma\,] \,,\\
\expv{\bar\psi_f\Gamma\psi_f}^{\rm LLL}_B
= \expv{\bar\psi_f P_f\Gamma P_f\psi_f}_B 
&= \frac{T}{4V}\frac{1}{\Z(0)} \int \!\D A_\mu \,e^{-S_g} \,{\det}^{1/4} [M(0)]\; \tr[M_f^{-1}(B)P_f\Gamma P_f\,] \,.
\end{split}
\label{eq:exppsiLLL}
\ee
The traces are evaluated using noisy estimators $\xi_j$. For the LLL projected 
observable, this amounts to
\be
\tr (M_f^{-1} P_f\Gamma P_f) = \tr (P_f M_f^{-1} P_f\Gamma P_f) =
\frac{1}{N_\xi}\sum_{j=1}^{N_\xi} \xi^\dagger_j P_f M_f^{-1} P_f
\Gamma P_f \,\xi_j\,, 
\ee
where $P_f$ is taken from Eq.~(\ref{eq:projdef3}) for the flavor $f$ 
and we used $P_f^2=P_f$ and the cyclicity of the trace.  
In the following we consider the quark condensate $\Gamma=\mathbb{1}$ and 
the spin polarization $\Gamma=\sigma_{xy}$ and refer to these by the
superscripts $S$ and  
$T$ (for scalar and tensor, respectively). The staggered discretization of the 
spin operator $\sigma_{xy}$ involves gauge links lying in the $x-y$ plane and 
is detailed in Ref.~\cite{Bali:2012jv}. 

Besides the representation of the traces using noisy estimators (which we use 
below to determine the observables), it is instructive to discuss
their relation to the overlap $W_i(\phi_k)$ of the four dimensional modes $\phi_k$ with the basis modes $\psi_{izt}$ carrying the two dimensional index $i$, defined in Eq.~\eqref{eq:overlapdef} and visualized in Fig.~\ref{fig:overlap}. 
To see this relation, we use the eigenmode basis of the four-dimensional Dirac operator
$\slashed{D}\phi_k=i\lambda_k \phi_k$. 
In this representation, the LLL-projected bilinears read
\be
 \begin{split}
  \tr[M_f^{-1}(B)P_f]
  &= 
   \sum_{k}\frac{m}{\lambda_k^2(B)+m^2}
   \sum_{i\leq N_c N_b}W_i(\phi_k)\,,\\
  \tr[M_f^{-1}(B)P_f\sigma_{xy}P_f]
  &\simeq 
   \sum_{k}\frac{m}{\lambda^2_k(B)+m^2}
   \sum_{i\leq N_c N_b} \sum_{\rm doublers} \sum_{z,t} 
   |\psi_{izt}^\dagger \phi_k|^2
   \cdot \overbrace{\psi_{izt}^\dagger\sigma_{xy}\psi_{izt}}^{{\varphi^{(z,t)}_i}^\dagger \sigma_{xy}^{(z,t)}\varphi^{(z,t)}_i}\\
   &\simeq \sum_{k}\frac{m}{\lambda^2_k(B)+m^2}
   \sum_{i\leq N_c N_b} W_i(\phi_k)\,
   \cdot \varphi_i^\dagger \sigma_{xy}\varphi_i\,,
 \end{split}
 \label{eq:tracesrep}
\ee
where we used the symmetry of $\slashed{D}$ that its eigenvalues appear in complex 
conjugate pairs. The spin operator $\sigma_{xy}$ 
is diagonal in the $z$ and $t$ coordinates, which allowed us to rewrite the matrix element 
in the second relation using the two-dimensional modes $\varphi_i$ and the block $\sigma_{xy}^{(z,t)}$ living on the slice $z,t$. 
In the first step of the second relation we used the fact that 
$\sigma_{xy}$ is to a good approximation diagonal in the two dimensional modes even in the presence of color interactions, see the right panel of Fig.~\ref{fig:spinxy}. Moreover, 
in the second step we approximated the matrix element of $\sigma_{xy}^{(z,t)}$ 
on the two-dimensional modes $\varphi_i^{(z,t)}$ to be 
independent of the coordinates $z,t$, which we find to hold if the average over gluon
configurations is performed.

In contrast to the LLL-projected observables of Eq.~(\ref{eq:tracesrep}), 
the full observables involve a sum over all values of $i$. 
The left panel of Fig.~\ref{fig:overlap} tells us that the contribution of the 
overlaps $W_{i\leq N_c N_b}(\phi_k)$ to the total $\sum_i W_i(\phi_k)$ is enhanced 
for low-lying modes $\phi_k$. 
Naively, this works in favor of the LLL dominance of the condensate, however, 
$\expv{\bar\psi_f\psi_f}$ also contains ultraviolet divergent contributions so that a sensible 
comparison of the LLL-projected and the full observables necessitates renormalization. 
The situation is similar for the spin polarization. In addition to the overlaps $W_i$, here the matrix elements $\varphi_i^\dagger\sigma_{xy}\varphi_i$ for $i\le N_c N_b$ are also much larger than for higher $i$, see the right panel of Fig.~\ref{fig:spinxy}, which (again, naively)
enhances the LLL-dominance for this observable even further. 
Our next step is therefore the renormalization of both observables, which we discuss
in the next subsection.

\subsection{Renormalization}
\label{sec:renormalization}

Both $\expv{\bar\psi_f\psi_f}_B$ and $\expv{\bar\psi_f\sigma_{xy}\psi_f}_B$ 
contain additive as well as multiplicative divergences.
However, it turns out that 
somewhat different renormalization procedures are required for the condensate and for
the spin polarization. 

Let us consider $\expv{\bar\psi_f\psi_f}_B$ first.
As the analytic 
calculation in the free case reveals (see App.~\ref{app:2}), the LLL
projected and the full condensates 
contain different divergences (logarithmic for the former and logarithmic plus 
quadratic divergences in the cutoff for the latter).
Thus, simply taking the difference of the two quantities is not
sufficient to cancel these terms. 
Instead, we consider two different routes to deal with these divergences. 

First, we use the gradient flow of the 
gauge fields to make both the LLL projected and the full observables
ultraviolet finite. This procedure smears the
gluon~\cite{Luscher:2010iy} and the fermion~\cite{Luscher:2013cpa}
fields over a smearing range $R_s$ and thereby 
eliminates ultraviolet noise and with that the additive divergent contribution to physical quantities.
The smearing radius is in spirit similar to a momentum cutoff $\Lambda=1/R_s$. 
We choose the magnetic field to set the 
smearing range: $R_s=c\cdot (eB)^{-1/2}$ with $c\approx 1$ and check 
that the results depend only mildly on $c$. 
Our implementation of the gluonic and fermionic flow for staggered 
quarks is detailed in Ref.~\cite{Bali:2014vja}. 
The so renormalized observable reads
\be
C^{S}_f \equiv
\left.\frac{\expv{\bar\psi_f\psi_f}^{\rm LLL}_B(R_s)}{\expv{\bar\psi_f\psi_f}_B(R_s)}\right|_{R_s=c/\sqrt{eB}}\,. 
\label{eq:CSf}
\ee

The second approach does not involve additional ultraviolet cutoffs like the 
smearing radius above. Instead, the additive divergences are canceled here 
by taking the difference between the expectation values at $B>0$ and at $B=0$. 
For the full condensate this is a straightforward procedure that gives the 
change of the condensate due to the magnetic field,
\be
\Delta\! \expv{\bar\psi_f\psi_f}_B =
\expv{\bar\psi_f\psi_f}_B - \expv{\bar\psi_f\psi_f}_{B=0}\,. 
\label{eq:subfullcond}
\ee
For the LLL projected condensate it is somewhat less obvious how to define this 
difference. 
The analysis of the free case (see App.~\ref{app:1}) reveals that the 
divergences can be canceled if one performs a similar projection in
the $B=0$ term as well, which involves the projector $\widetilde{P}$ of
Eq.~(\ref{eq:projdef_extra}), built from the eigenmodes of the $B=0$ Dirac operator.
We then define 
the subtracted LLL condensate as
\be
\Delta\! \expv{\bar\psi_f\psi_f}^{\rm LLL}_B =
\expv{\bar\psi_f\psi_f}^{\rm LLL}_{B} -
\big\langle\bar\psi_f\widetilde{P}_f\psi_f\big\rangle_{B=0}\,. 
\ee
We emphasize that in the $B>0$ term the projector projects on the $N_cN_b\times 2$ 
lowest eigenmodes of the two-dimensional $B>0$ Dirac operator. 
In the
$B=0$ term, $\tilde{P}_f$ projects on the same number of modes, but
this time of the $B=0$ Dirac operator.
Using this construction, our second ratio reads\footnote{Notice that $D^S_f$ involves the $B=0$ projector $\widetilde{P}$, which -- according to the discussion at the end of Sec.~\ref{sec:fourd} -- corresponds to a nonlocal operator and might complicate the 
continuum limit of $D^S_f$. 
From this point of view, the first ratio $C^S_f$ is more advantageous. Nevertheless, below in Sec.~\ref{sec:results} we will find that both $C^S_f$ and $D^S_f$ give similar results.}
\be
D^{S}_f \equiv \frac{\Delta\! \expv{\bar\psi_f\psi_f}^{\rm LLL}_B}{\Delta\! \expv{\bar\psi_f\psi_f}_B}\,.
\label{eq:DSf}
\ee

For the spin polarization (superscript $T$), the ratio $C_f$ can be defined using the same prescription as 
for the condensate,
\be
C^{T}_f \equiv
\left.\frac{\expv{\bar\psi_f\sigma_{xy}\psi_f}^{\rm LLL}_B(R_s)}{\expv{\bar\psi_f\sigma_{xy}\psi_f}_B(R_s)}\right|_{R_s=c/\sqrt{eB}}\,. 
\label{eq:CTf}
\ee
The ratio $D_f^T$ must be defined differently, since
$\expv{\bar\psi_f\sigma_{xy}\psi_f}_B$ vanishes identically at
$B=0$. Namely, in the 
absence of the magnetic field, there is no preferred direction and the spin 
polarization averages to zero. However, we can exploit the fact that the 
divergences of the LLL and of the full observable coincide this time. 
This is supported by the calculation in the free case in App.~\ref{app:2}. 
This divergent piece, denoted by $T^{\rm div}_f$, has been
determined\footnote{In fact, the determination of $T_f^{\rm div}$ in
  Ref.~\cite{Bali:2012jv} was carried out with the magnetic field both
  in the valence and in the sea sector taken into account. This we checked to 
  be a sub-percent effect compared to $\expv{\bar\psi_f\sigma_{xy}\psi_f}_B$ at low 
  temperatures, but becomes increasingly important as $T$ grows and the expectation 
  value reduces. We find that the systematic error in $T^{\rm div}$ due to neglecting 
  the sea contribution is much smaller than lattice artefacts for our lowest two temperatures 
  $T=124 \textmd{ MeV}$ and $T=170\textmd{ MeV}$ so in the following we only 
  consider these simulation points for $D^T_f$.} at zero temperature in Ref.~\cite{Bali:2012jv} using the method
summarized in App.~\ref{app:2}. Therefore we have
\be
D^{T}_f \equiv 
	\frac{
  \expv{\bar\psi_f\sigma_{xy}\psi_f}^{\rm LLL}_B - T_f^{\rm div}
  }{\expv{\bar\psi_f\sigma_{xy}\psi_f}_B-T_f^{\rm
    div}}\,. 
    \label{eq:DTf}
\ee

Let us now turn to the multiplicative renormalization. 
The renormalization constants are expexted to be 
independent of the magnetic field, and the LLL-approximation is
assumed to accurately describe  
strong magnetic fields. Thus it seems natural to assume that the
renormalization constants in the full theory and for the LLL coincide,
and ratios of the LLL-projected  
and the full observables -- like $C_f$ and $D_f$ above -- are free of
multiplicative divergences.  
However, since defining the LLL-projection on a finite lattice 
effectively involves the asymptotic limit $B\to\infty$ {\it before} 
the continuum limit $a\to0$ -- i.e.\ with 
the magnetic field exceeding even the cutoff -- the ultraviolet
behavior might still 
be affected. Whether this is the case should be checked in the future. 
For this reason, in the present paper we do not perform a continuum 
extrapolation of our results but merely show data obtained using
different cutoffs.

\section{Results}
\label{sec:results}

We have performed measurements for a range of temperatures and magnetic fields 
using four lattice ensembles with $N_t=6,8,10$ and $12$. These serve 
to approach the continuum limit $a\to 0$ at a fixed temperature $T$
owing to $T=1/(N_t a)$. Thus, the larger $N_t$, the closer we are to the continuum.
The aspect ratios were set to $N_s/N_t=4$ to keep the physical volume fixed. 
Throughout the rest of this section 
we consider the down quark flavor $f=d$. Since we only implement the magnetic field 
and the LLL-projector in the valence sector, for the bilinears for the up quark we 
identically have $\expv{\bar\psi_u\Gamma\psi_u}_B = \expv{\bar\psi_d\Gamma\psi_d}_{2B}$ 
(here we also exploited parity symmetry $B\leftrightarrow -B$). 

\begin{figure}[ht!]
 \centering
 \includegraphics[width=8cm]{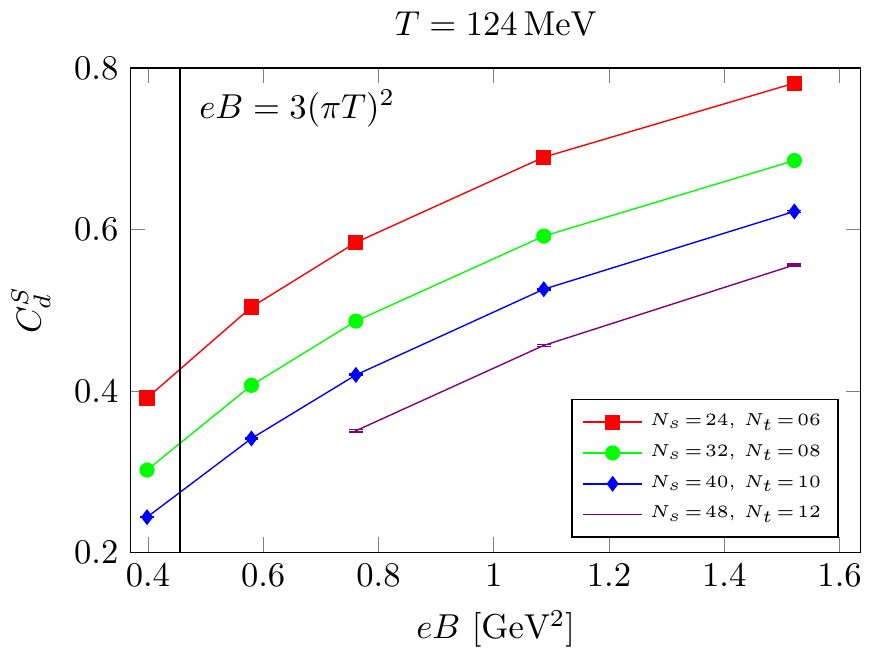} \hspace*{.4cm}
 \includegraphics[width=8cm]{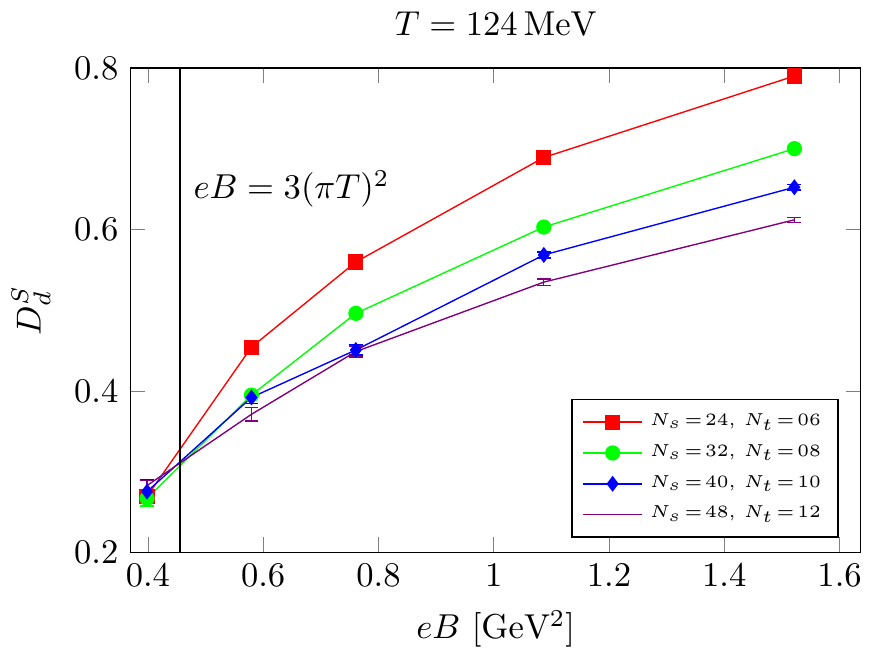}\\
 \hspace*{.3cm}\includegraphics[width=8.1cm]{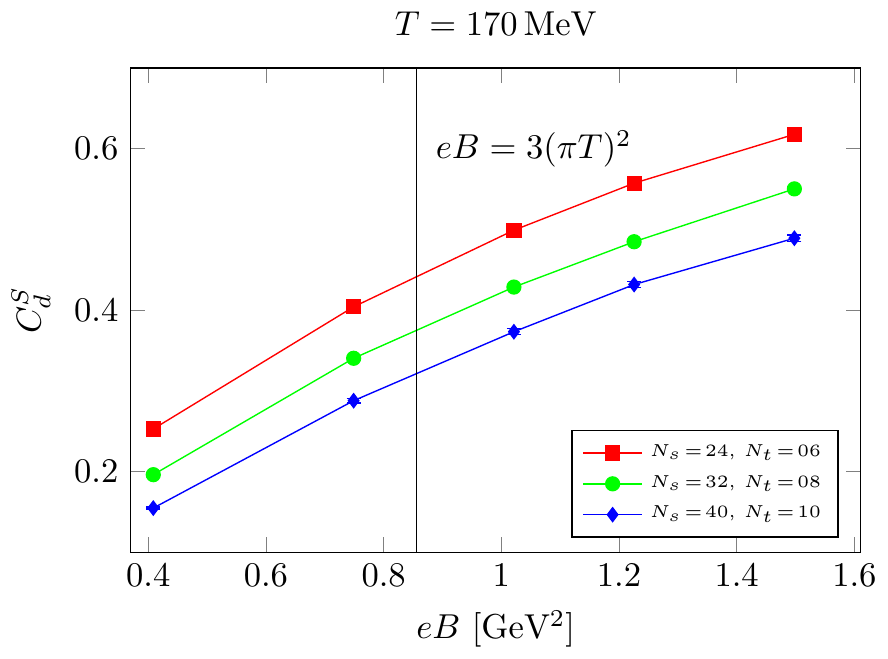} \hspace*{.2cm}
 \includegraphics[width=8.1cm]{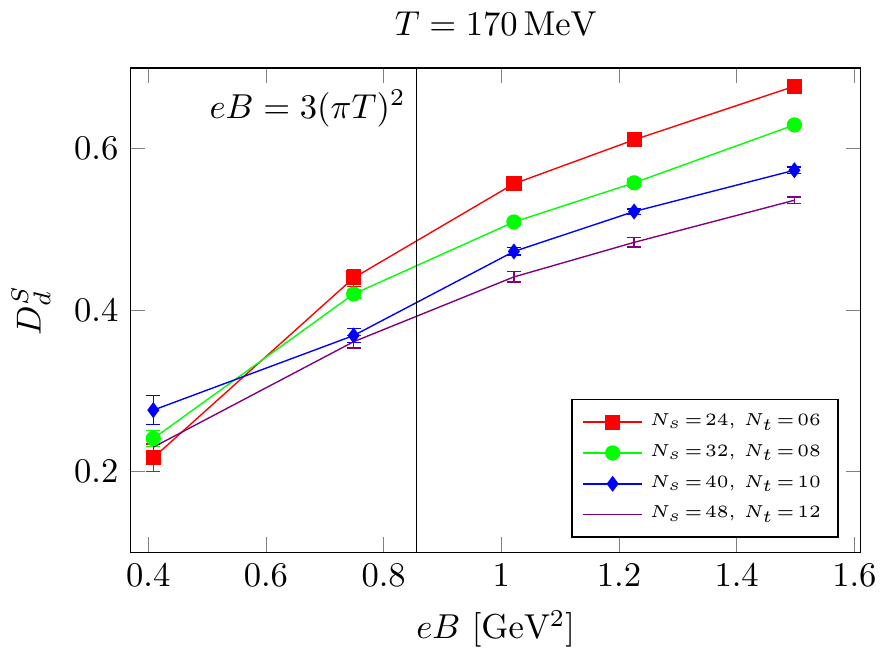}
 \caption{\label{fig:cds}The ratios $C_d^S$ (left panels) and 
 $D_d^S$ (right panels) for the quark condensate as functions of the magnetic field 
 at a temperature $T=124\textmd{ MeV}$ (upper panels) and $T=170\textmd{ MeV}$ (lower panels).}
\end{figure}

We begin the presentation of the results with the quark condensate. 
The two ratios $C_d^S$ and $D_d^S$ of Eqs.~(\ref{eq:CSf}) and~(\ref{eq:DSf}) 
are plotted in Fig.~\ref{fig:cds} for two temperatures, below and above the 
finite temperature QCD crossover. 
Both combinations give similar results, rising from 
about $30\%$ to $60\%-80\%$ within the range $0.4 \textmd{ GeV}^2 < eB < 1.5 \textmd{ GeV}^2$. 
The ratios are expected to approach unity for strong magnetic fields, as the analytic 
calculation in the free case shows (see App.~\ref{app:1}). 
In the figures we also include the $eB=3(\pi T)^2$ vertical line to indicate the point 
where the magnetic field (times the modulus of the electric charge $|q_d|=e/3$) 
becomes the largest dimensionful scale in the theory. 
We measured $C^S_d$ with smearing radii $R_s=1 \cdot (q_dB)^{-1/2}$ and
$R_s=2/\sqrt{3} \cdot (q_dB)^{-1/2}$ and found that the systematic effect due to varying $R_s$ 
in this range is much smaller than the lattice artefacts. 

While for the ratio renormalized through the gradient flow, our $N_t=10$ and 
$N_t=12$ results still significantly differ, $D_d^S$ shows nice scaling towards the 
continuum limit. As $B$ grows and the magnetic field in lattice units increases, lattice 
artefacts are expected to become more pronounced, as visible in the right panels of 
Fig.~\ref{fig:cds}. In the following we will take the $N_t=12$ results for $D_d^S$ as 
a reference for the validity of the LLL-approximation.
The LLL-contribution to this observable reaches 60\% at our largest magnetic field and seems to further increase towards unity rather slowly. This is in agreement with the free case, where the deviation from 1 is of the form $1/\log(B)$ for large $B$, see Eq.~\eqref{eq:freecase_approachtounity}.

\begin{figure}[ht!]
 \centering
 \includegraphics[width=8cm]{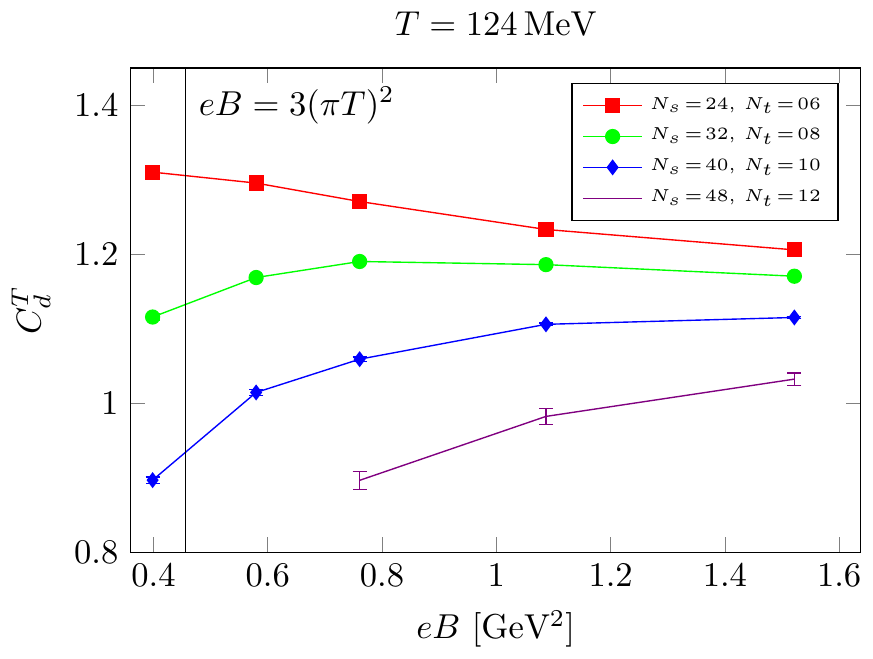} \hspace*{.4cm}
 \includegraphics[width=8cm]{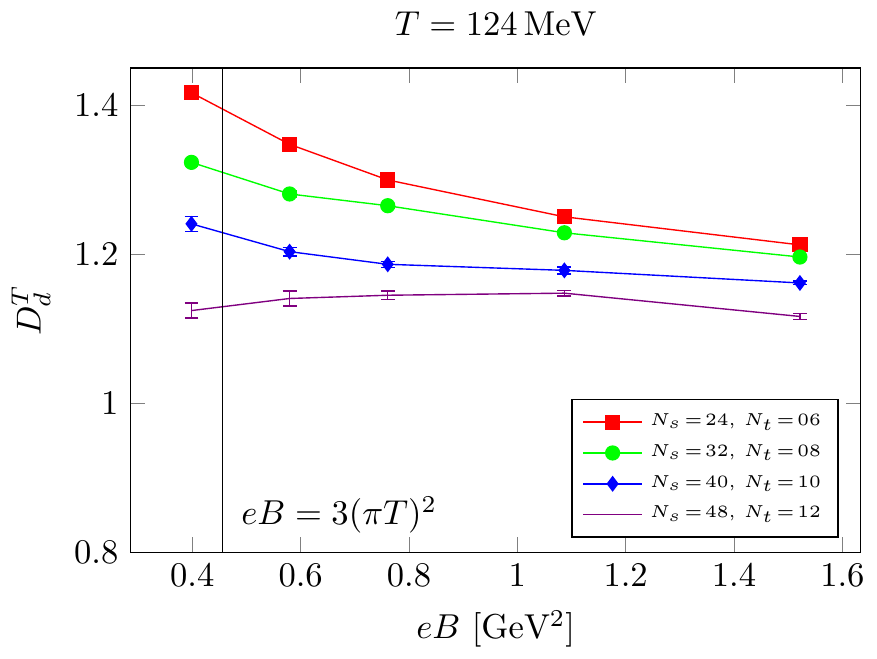}\\
 \hspace*{.3cm}\includegraphics[width=8.1cm]{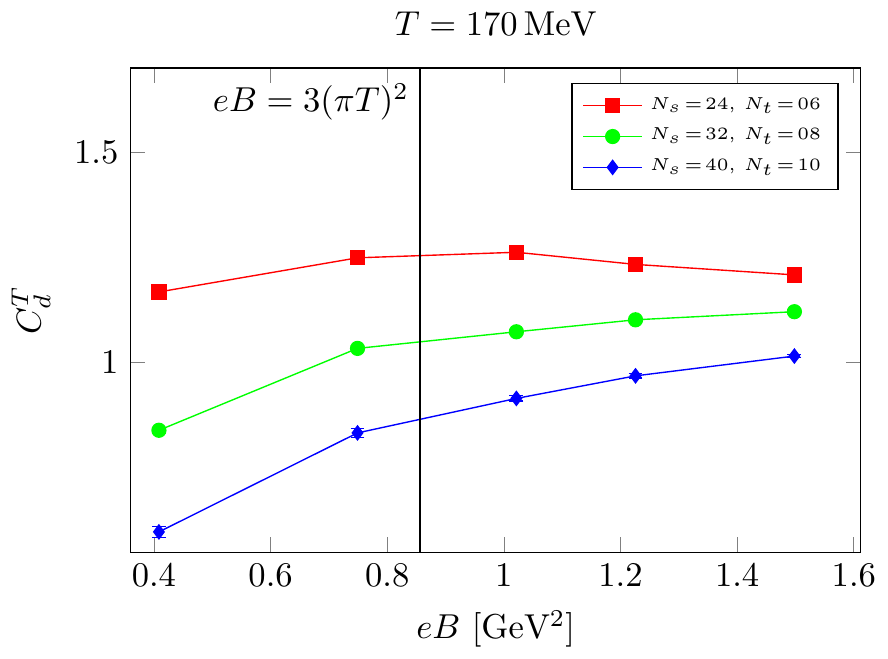} \hspace*{.2cm}
 \includegraphics[width=8.1cm]{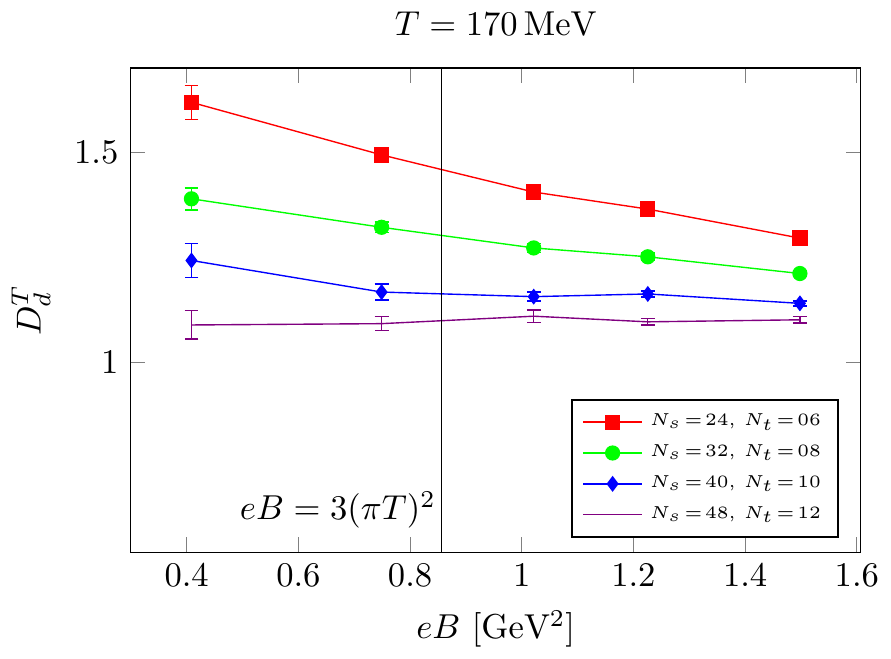}
 \caption{\label{fig:cdt}The ratios $C_d^T$ (left panels) and 
 $D_d^T$ (right panels) for the spin polarization as functions of the magnetic field 
 at a temperature $T=124\textmd{ MeV}$ (upper panels) and $T=170\textmd{ MeV}$ (lower panels).}
\end{figure}

Next we turn to the spin polarization. The renormalized ratios $C_f^T$ and $D_f^T$ were defined 
in Eqs.~(\ref{eq:CTf}) and~(\ref{eq:DTf}) above. We plot both in Fig.~\ref{fig:cdt} 
as functions of the magnetic field for the two temperatures that we considered above. 
Just as for 
the condensate, the ratio $D_d^T$ exhibits faster scaling towards the continuum limit.
We find that $D^T_d>1$ i.e., 
the spin polarization is overestimated by the LLL approximation (for $C^T_d$ this trend 
is not obvious due to large cutoff effects).
This may be understood by noting that $\sigma_{xy}$ is a traceless operator and that 
$\sigma_{xy}$ has matrix elements close to unity on the LLL-modes (see Fig.~\ref{fig:spinxy}). 
Thus, the higher modes must have negative matrix elements so that the total trace can vanish
and, accordingly, the HLL contribution to $\bar\psi\sigma_{xy}\psi$ is negative. 
The 
deviation of $D^T_d$ from unity is much milder than for the condensate, 
remaining below $15\%$ for our finest lattices for the 
complete range of magnetic fields that we consider here. Notice that the LLL-approximation 
is expected to work well for this observable since in the free case the HLL-contribution to 
$\bar\psi\sigma_{xy}\psi$ vanishes identically (see App.~\ref{app:1}). 

\section{Summary}
\label{sec:summary}

In this paper we investigated the validity of the lowest-Landau-level
(LLL) approximation to QCD in the presence of background magnetic
fields. In the absence of color interactions, this approximation is
based on the structure of the analytically calculable energy levels
(the Landau levels) of the quantum system. While the energy of the  
lowest level is independent of $B$, the higher levels have squared energies
above $B$ and thus become negligible if the magnetic field is
sufficiently strong. 
Furthermore, the characteristic degeneracy of the Landau levels is proportional to the magnetic flux.

The presence of (nonperturbative) color interactions mixes the levels
and therefore complicates this simple picture considerably. In the
present paper we demonstrated, for the first time, that the lowest  
Landau level can nevertheless be defined in a consistent manner even
for strongly interacting quarks. The definition of the LLL is based on
a two-dimensional topological argument that characterizes the $x-y$
plane (the plane perpendicular to the magnetic field).  
Namely, the two-dimensional LLL modes have zero energy and their number is 
a topological invariant fixed by the flux of 
the magnetic field, independently of the gluonic field configuration. 
Although these exact zero modes are shifted to nonzero values on a
finite lattice, they are still well separated from the rest by a gap
in the spectrum. We have shown that this gap is a remnant of the
largest gap in the fractal structure usually referred to as
Hofstadter's butterfly in the Hofstadter (lattice) model of solid state
physics. 

This construction can be performed on each $x-y$ plane, i.e., for each
value of the $z$ and $t$ spacetime-coordinates. 
While the two-dimensional modes can be unambiguously classified as 
belonging or not to the LLL, in four dimensions this is not the case anymore: 
a general four-dimensional Dirac eigenmode has overlap both in and out
of the LLL.  
We defined the projector $P$ that projects the four-dimensional modes onto the 
subspace of two-dimensional LLL modes for each $z$ and $t$.
Using the projector, 
we have shown that low-lying four dimensional modes have enhanced overlap with the LLL. 
For higher four-dimensional modes the overlap with LLL is instead suppressed with respect 
to HLLs.

Motivated by this, the LLL contribution to standard fermionic
observables can be determined. In particular, we concentrated on the
quark condensate $\bar \psi P \psi$ and the spin polarization
$\bar\psi P \sigma_{xy} P \psi$ for the down quark. We constructed
ratios of the LLL-projected and the full observables that are free of
additive divergences (demonstrated in the free case in
App.~\ref{app:2}) and that approach unity in the $B\to\infty$ limit  
(shown in the free case in App.~\ref{app:1}).

\begin{figure}[ht!]
 \centering
 \includegraphics[width=8cm]{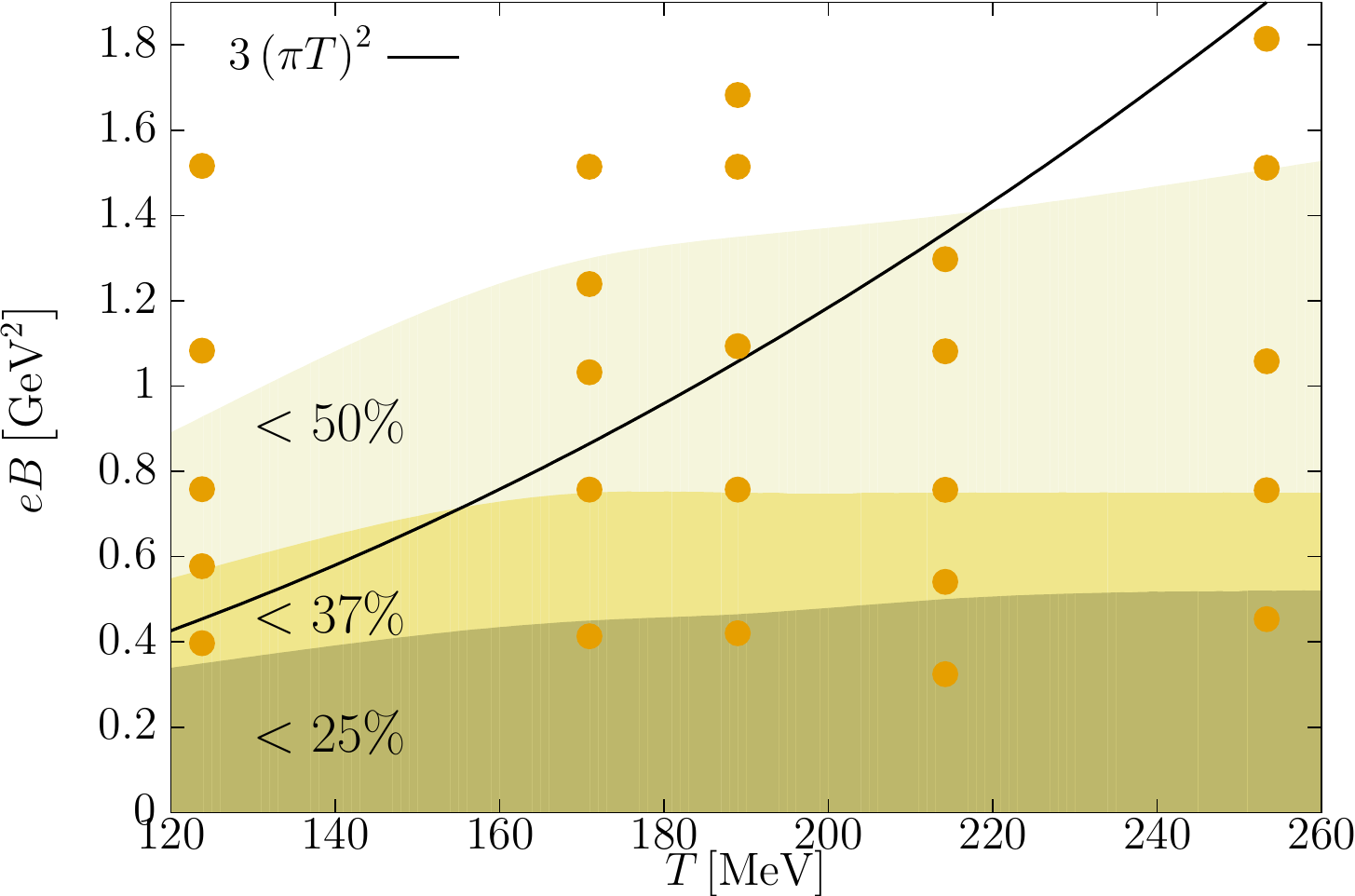}
 \caption{\label{fig:summ}Visualization of the validity of the LLL-approximation 
 for the down quark condensate. The lighter the color, the closer the LLL-projected 
 condensate is to the full result. The orange dots denote our simulation points and the 
 solid black line marks $q_dB=(\pi T)^2$.}
\end{figure}

Our results indicate that the LLL approximation
underestimates the quark condensate and overestimates 
the spin polarization. In addition, the LLL-projected quantities
slowly approach the full observables as 
the magnetic field grows and exceeds further dimensionful scales
$\Lambda_{\rm QCD}^2$ and $(\pi T)^2$ in the system. 
Our final results for the condensate (using the $N_t=12$ data for $D^S_d$) are visualized 
for a wide range of temperatures 
in Fig.~\ref{fig:summ}. In this figure the validity of the LLL-approximation is 
represented in the $B-T$ plane. Dark colors stand for regions where the approximation 
breaks down in the sense that the LLL-projected condensate is far away from its 
full value (so that $D^S_d$ is less then $25\%$, $37\%$ or $50\%$). The white region is 
where the LLL-contribution to the condensate amounts to more than half of the full condensate. 
The contours were determined by means of a spline interpolation of $D^S_d(B)$ to calculate the 
magnetic fields where the observable reaches a given percent. Using these 
threshold magnetic fields for each temperature, a second set of spline interpolations results in 
continuous $T$-dependent functions that are shown in the contour plot. The so obtained 
contours may be compared to the naive expectation 
$q_dB\gtrless (\pi T)^2$, also indicated in the figure.

Summarizing, we have quantified the systematics of the LLL approximation 
via first-principle lattice simulations of QCD with background magnetic fields. 
The results may be compared directly to low-energy models or effective
theories employing only the lowest Landau level.
We emphasize that our findings correspond to the valence sector, i.e.\
only the quark fields in the operators $\bar\psi \psi$ and
$\bar\psi\sigma_{xy}\psi$ are projected to the LLL but not the virtual
sea quarks appearing in loops. The extension of the approach to sea
quarks is more involved and is left for a future study. 

\acknowledgments

This research was supported by the DFG (Emmy Noether Programme EN 1064/2-
1, SFB/TRR 55, BR 2872/6-1, BR 2872/6-2 and BR 2872/7-1), by OTKA (OTKA-K-113034) and 
by the Hungarian Academy of Sciences under ``Lend{\"u}let'' (No.\ LP2011-011).
GE is grateful for the hospitality of the organizers of the Workshop on Magnetic Fields in Hadron Physics 2016, 
where early results of our work were presented and discussed.

\appendix

\section{Additive divergences in the free case}
\label{app:2}

In this appendix we calculate the additive divergences of the fermion bilinears 
in the free case and demonstrate that the ratios $C_f$ and $D_f$ of Eqs.~(\ref{eq:CSf}), (\ref{eq:DSf}), (\ref{eq:CTf}) and~(\ref{eq:DTf})
are ultraviolet 
finite. We work in a finite but large volume $V=L^3$, orient the magnetic field in the positive
$z$ direction and assume that $qB>0$. Since the divergences are independent of the temperature,
we will work at $T=0$. 
Throughout the appendices we will neglect a
factor $N_c=3$, since in the free case all $N_c$ colors give the same
contribution. 

The quark condensate in the free case (four dimensions, no LLL
projection yet) can easily be shown to  
have the following spectral representation (see e.g.\ App.~B in \cite{Bali:2012jv}):
\be
\expv{\bar\psi\psi}_B
 =\frac{T}{V}\,\tr (\slashed{D}+m)^{-1}
 =\sumint_{\,\lambda} 
 \frac{1}{i\lambda+m}
 =\sumint_{\,\lambda>0} 
 \frac{2m}{\lambda^2+m^2}, \quad\quad
 \sumint_{\,\lambda>0}=T\sum_{p_t}\sum_n\int_{-\infty}^\infty \!
\frac{dp_z}{2\pi}\, \frac{\nu_{n p_z p_t}}{L^2}\,,
 \label{eq:PBP_spectral}
\ee
where in the third equality we used the existence of chiral partners
with opposite eigenvalue $\lambda$. The eigenvalues and the degeneracies are given in
Eq.~(\ref{eq:freecase4d}). 
For $T=0$ the sum over Matsubara frequencies turns into 
an analogous integral over $p_t$. The combined momentum
integration/summation over $p_z$ and $p_t$ 
is UV divergent for fixed
$n$. Therefore, the condensate is divergent for the LLL projected as
well as for the full case. 

We can make this divergence more transparent with the help of
Schwinger's proper time~\cite{Schwinger:1951nm}. In our case this simply amounts to using the
identity  $1/y=\int_0^\infty\!ds\,\exp(-ys)$ to exponentiate
$\lambda^2+m^2$ making it possible to sum over the Landau levels and
to integrate over the momenta, 
\be
\label{eq:PBP_FULL}
\begin{split}
\expv{ \bar\psi\psi}_B
 &=\,\frac{qB m}{2\pi^2}
 \int_0^\infty\!ds\,\exp(-m^2s)
  \sum_{n}(2-\delta_{n,0})\exp(-2qBns)
 \int_0^\infty\!dp\, p\exp(-p^2s)\\
 &=\,\frac{qB m}{4\pi^2}
 \int_0^\infty\!ds\,\exp(-m^2s)\coth(qBs)\,\frac{1}{s}\,.
\end{split}
\ee
The quantity $s$ has dimension $1/m^2$ and thus the UV divergence
occurs at the lower end of the integral, where 
\be
\expv{ \bar\psi\psi}_B=
 \frac{qB m}{4\pi^2}\int_{1/\Lambda^2}\!ds\,\exp(-m^2s)\,
 \left[\frac{1}{qBs^2}+\frac{qB}{3}+\mathcal{O}(s^0)\right]
 = \frac{m\Lambda^2}{4\pi^2}
 - \frac{m^3}{4\pi^2} \log\frac{\Lambda^2}{m^2} 
 +\text{finite}\,.
\ee
Since both of the divergent terms are $B$-independent, the additive divergence in the condensate can be removed by subtracting the $B=0$ condensate.

The LLL projected condensate follows easily by setting $n=0$ above instead of summing over $n$, 
or, equivalently, by replacing $\coth(qBs)$ by 1. We obtain
\be\label{eq:PBP_LLL}
\expv{\bar\psi\psi}_B^{\text{LLL}}
 =\,\frac{qB m}{4\pi^2}
 \int_0^\infty\!ds\,\exp(-m^2s)\,\frac{1}{s}\,.
\ee
Consequently, the divergence is weaker:
\be
\expv{\bar\psi\psi}_B^{\text{LLL}}=
  \frac{qBm}{4\pi^2} \log \frac{\Lambda^2}{m^2} 
 + \text{finite} \,.
 \label{eq:pbpLLLfree}
\ee 
The magnetic field occurs only as an overall factor, and the LLL
projected condensate would vanish for $B=0$, where indeed the notion
of Landau levels is meaningless. 
The additive divergence can be canceled by subtracting the fermion bilinear 
involving the projector $\widetilde{P}$ defined at $B=0$, see App.~\ref{app:1} below.

Another way of regularizing the observables is the gradient flow
method \cite{Luscher:2010iy,Luscher:2013cpa}.  
There the fields are flowed/smeared with the help of the heat kernel 
\be
 K_t=e^{-t(-D^2+m^2)}\,,
\ee
where $D^2$ is the gauge-covariant Laplace operator and $t$
is the flow time, of dimension $1/m^2$. Note that $-D^2$ is
nonnegative.  In the condensate, two quark fields are flowed, thus 
\be
\expv{ \bar\psi\psi}_B(t)
 =\frac{T}{V}\,\tr [(\slashed{D}+m)^{-1}K_{2t}]\,.
\ee
Note that the argument $t$ is related to the smearing radius $R_s$
introduced in Eq.~(\ref{eq:CSf}) in the main text as $t=R_s^2/8$~\cite{Luscher:2010iy}. 

For the evaluation of the free flowed condensate we note that the free
Laplacian commutes with $\slashed{D}^2$ and its eigenvalues are those
of the latter, Eq.~(\ref{eq:freecase}) and (\ref{eq:freecase4d}), with
the spin $s_z$ set to zero, i.e., with quantum number $l$ and the
corresponding degeneracy $\nu_l$ 
\be
 \text{eigenvalues}(-D^2+m^2)
 =qB(2l+1)+p_z^2+p_t^2+m^2~,\quad
 l\in\mathbb{Z}_0^+~,\quad
 \nu_l=N_b \,.
\ee
The contributions of the chiral partners to the condensate can be taken into account as before, and thus we obtain the spectral representation
\be
\begin{split}
 \expv{\bar\psi\psi}_B(t)
 &=\frac{ qB}{2\pi^2}\,m\, e^{-2t m^2}
 \sum_{l=0}^\infty\sum_{s_z=\pm 1/2}
 \int_0^\infty \!dp\, p\,\frac{\exp(-2t\{qB(2l+1)+p^2\})}{qB(2l+1-2s_z)+p^2+m^2}\,.
\end{split}
\ee
Also here we can make use of Schwinger's proper time, which yields
\be
 \label{eq:PBP_FULL_flowed}
\expv{ \bar\psi\psi}_B(t)
 =
 \frac{qB}{4\pi^2} 
 m\, e^{-2tm^2}
 \int_0^\infty ds\, 
 \frac{e^{-sm^2}\cosh(qBs)}{(2t+s)\, \sinh(qB(2t+s))}\,. 
\ee
Clearly, the flow time $t$ regularizes the integral for small values of the proper time $s$,
and is hence equivalent to a UV-cutoff in momentum space or a smearing radius in coordinate space. 
In the LLL approximation the flowed condensate is obtained by taking only the $l=0, s_z=+1/2$ contribution 
to the full condensate:
\be
 \label{eq:PBP_LLL_flowed}
 \expv{\bar\psi\psi}_B^{\rm LLL}(t)
 =
 \frac{qB}{4\pi^2} 
 m\, e^{-2tm^2} e^{-2tqB}
 \int_0^\infty ds\, 
 \frac{e^{-sm^2}}{(2t+s)}\,. 
\ee
For the LLL approximation to make sense, $B$ should be the largest
scale in the system.  
Therefore we choose for the flow time $t = c^2/(8qB)$ or, equivalently, 
for the smearing radius $R_s=c/\sqrt{qB}$, with $c\approx 1$. 
Note that this choice has a well defined continuum limit for fixed $B$, 
as the physical smearing radius is kept fixed by $B$. 

The other observable we consider is the spin polarization 
$\bar\psi \sigma_{xy}\psi$. Following
Eq.~\eqref{eq:PBP_spectral} we obtain
\begin{align}
	\expv{\bar\psi \sigma_{xy}\psi}_B
	&=
 \sumint_{\,\lambda>0}
 \frac{2m}{\lambda^2+m^2}
	\langle \lambda | \sigma_{xy} |\lambda \rangle
	=
	\expv{\bar\psi \sigma_{xy}\psi}_B^{\rm LLL}
	=
	\expv{\bar \psi\psi}_B^{\rm LLL}\,,
\end{align}
where in the last two equalities we have used that the expectation value, 
$\langle \lambda^{\rm LLL} | \sigma_{xy} |\lambda^{\rm LLL} \rangle = 1$,
is only non-vanishing in the lowest Landau level. 
Thus in the free case the spin polarization is made up purely by the LLL-contribution. 
Both are equal to Eq.~(\ref{eq:pbpLLLfree}) and are thus logarithmically divergent.
Since the divergence is proportional to $m\log(\Lambda^2/m^2)$, a possible 
way to separate the infinite part is via~\cite{Bali:2012jv}
\be
	T^{\rm div}
	\equiv
	m\,\frac{\partial}{\partial m}\, \expv{\bar\psi \sigma_{xy}\psi}_B, \quad\quad
	\expv{\bar\psi\sigma_{xy}\psi}_B - T^{\rm div} = \frac{mqB}{2\pi^2} = \textmd{finite}\,.
\ee
This prescription can also be applied in the interacting case and $m\frac{\partial}{\partial m}\, \expv{\bar\psi\sigma_{xy}\psi}_B$ has been measured in full QCD in Ref.~\cite{Bali:2012jv}.

\section{\boldmath $B$-dependence of the ratios in the free case}
\label{app:1}

In this appendix we discuss the magnetic field-dependence of the ratios $C$ and $D$
in the free case. We again assume that $qB>0$ and neglect an overall factor of $N_c$.  
First we consider the ratio $D^S$ of Eq.~(\ref{eq:DSf}) for the quark condensate,
\be
D^S(B)\equiv \frac{\Delta\expv{\bar\psi\psi}_B^{\rm LLL}}{\Delta
  \expv{\bar\psi\psi}_B} 
= \frac{\expv{\bar\psi\psi}^{\rm LLL}_B - \expv{\bar\psi\widetilde{P}\psi}_{B=0}}{\expv{\bar\psi\psi}_B-\expv{\bar\psi\psi}_{B=0}}
\ee
where the projector $\widetilde{P}$, defined in Eq.~(\ref{eq:projdef_extra}), 
involves the lowest $N_b$ two-dimensional
modes (for each $z-t$ slice), with $N_b$ given by the flux quantum 
that corresponds to the finite $B$ term. 
We discuss here the $T=0$ case in detail.

In the free case the denominator reads, cf.\ Eqs.~\eqref{eq:PBP_FULL} and \eqref{eq:PBP_LLL} with variable change $s\to s/qB$,
\be
\begin{split}
\Delta\expv{\bar\psi\psi}_B &=
\frac{m qB}{4\pi^2} \int \frac{\dd s}{s^2} \left( s \coth s -1 \right)
e^{-m^2s/qB}\\
&= \frac{m qB}{2\pi^2} \left[ \log \Gamma(x) -\left(x-\frac{1}{2}\right) \log x + x
-\frac{1}{2} \log(2\pi) \right]\,,
\end{split}
\label{eq:PBP_FULL_integrated}
\ee
where $x=m^2/(2qB)$. The numerator is the difference
\be
\begin{split}
\Delta \expv{\bar\psi\psi}^{\rm LLL}_B
&= \frac{qB}{2\pi} \int \frac{\dd p_{zt} p_{zt}}{2\pi} \frac{2m}{p_{zt}^2+m^2}
- 2
\int_0^\infty \frac{\dd p_{zt} p_{zt}}{2\pi} 
\int_0^{\sqrt{qB}} \frac{\dd p_{xy} p_{xy}}{2\pi} 
\frac{2m}{p_{xy}^2+p_{zt}^2+m^2} \\
&= 
\frac{m}{\pi}\int_0^\infty \frac{\dd p_{zt} p_{zt}}{2\pi} 
\left[ \frac{qB}{p_{zt}^2+m^2} -  2
\int_0^{\sqrt{qB}} \frac{\dd p_{xy} p_{xy}}{p_{xy}^2+p_{zt}^2+m^2} 
 \right]
\end{split}
\ee
which turns out to be UV finite. 
Here we subtract the $B=0$ contribution of the 
lowest $N_b$ modes from the LLL contribution. 
To see why $\sqrt{qB}$ is the 
correct value for the upper limit, consider the number of two-dimensional 
fermionic quantum states at $B=0$ in a finite box $L_xL_y$,
\be
2 L_x L_y \int_0^\xi\frac{\dd p p}{2\pi} = N_b, \quad\to\quad \xi =
\sqrt{\frac{2\pi N_b}{L_xL_y}} = \sqrt{qB}\,.
\ee
Performing the integrals we obtain
\be
\Delta\expv{ \bar\psi\psi}^{\rm LLL}_B
= \frac{m qB}{4\pi^2} 
\left[ \left(1+2x\right)\log\left(1+\frac{1}{2x}\right) -1 \right]
\ee
and with Eq.~\eqref{eq:PBP_FULL_integrated}:
\be
D^S(B)\xrightarrow{B\to\infty}
1 + \mathcal{O}\left(\frac{1}{\log (qB/m^2)}\right)\,.
\label{eq:freecase_approachtounity}
\ee

Secondly we consider the flowed ratio $C^S(B)$, i.e., 
\be
	C^S(B) 
	=
	\frac{\expv{\bar\psi\psi}^{\rm
            LLL}_B(t)}{\expv{\bar\psi\psi}_B(t)} ,  
\ee
where the flow time is set by the magnetic field, $t=c^2/(8qB)$, 
and $c\approx 1$ is a fixed parameter. 
$C^S(B)$ can be represented as the ratio of two integrals, cf. \eqref{eq:PBP_FULL_flowed} and \eqref{eq:PBP_LLL_flowed}, $I_1$ and $I_2$: 
\be
	C^S(B)
	=
	\frac{I_2(t)}{I_1(B,t)} 
\ee
with
\be
	I_1 
	=
	\int_0^\infty ds\, \frac{e^{-sm^2}}{2t+s} \frac{1+e^{-2qBs}}{1-e^{-2qB(2t+s)}},
	\hspace{8em}
	I_2 
	=
	\int_0^\infty ds\, \frac{e^{-sm^2}}{2t+s}.
\ee
In the limit $B\to \infty$ $I_2$ has the asymptotic form
\be
	I_2
	\xrightarrow{B\to\infty}
	\log(qB/2cm^2). 
\ee
For $C^S(B)$ we get
\be
	\frac{1}{C^S(B)} 
	=
	1 + \frac{I_1 - I_2}{I_2}
	\xrightarrow{B\to\infty}
	1 + {\O\left( \frac{1}{\log(qB/m^2)} \right)}, 
\ee
since $I_1 - I_2$ is bounded by
\be
  0
	\le
	I_1 - I_2 
	\le
	\frac{\coth(2c^2)}{4c^2}.
\ee
This inequality comes about by noting that
\be
\begin{split}
	I_1 - I_2
	&=
	\int_0^\infty ds\,
	\frac{e^{-sm^2}e^{-2sqB}}{2t+s}  \frac{1+e^{-4qBt}}{1-e^{-4qBt}e^{-2qBs}} 
	\\
	&\le
	\frac{1+e^{-4qBt}}{1-e^{-4qBt}} 
	\int_0^\infty ds\,
	\frac{e^{-sm^2}e^{-2sqB}}{2t+s}  
	=
	\frac{1+e^{-4qBt}}{1-e^{-4qBt}} 
	\int_0^\infty ds\,
	\frac{e^{-sm^2/qB}e^{-2s}}{2qBt+s}  
	\\
	&\le
	\frac{1}{4qBt} 
	\frac{1+e^{-4qBt}}{1-e^{-4qBt}} .
\end{split}
\ee
Thus also the flowed ratio,
\be
C^S(B) \xrightarrow{B\to\infty}
1 + 
\mathcal{O}\left(\frac{1}{\log (qB/m^2)}\right)
\ee
becomes unity as expected for large $B$.

For nonzero temperature, $T\neq 0$, the calculation is somewhat more
involved. However, it turns out that additive divergences cancel in
$D^S(B)$ as in the $T=0$ case, and one finds
\begin{equation}
  \label{eq:finiteT}
  \begin{aligned}
     \Delta\expv{\bar\psi\psi}_B &=        
 \frac{m qB}{(2\pi)^2} 
\int_0^\infty \frac{ds}{s^2}\,
e^{-s \frac{m^2}{qB}} (s\coth s-1)\,
\bar{\Theta}_2\left(e^{-s \frac{(2\pi T)^2}{qB}}\right)
\,,\\
\Delta\expv{\bar\psi\psi}^{\rm LLL}_B
&= \frac{m qB}{(2\pi)^2} 
\int_0^\infty \frac{ds}{s^2}\,
e^{-s \frac{m^2}{qB}}(s-1 + e^{-s})\,
\bar{\Theta}_2\left(e^{-s \frac{(2\pi T)^2}{qB}}\right)
\,,
  \end{aligned}
\end{equation}
where
\begin{equation}
  \label{eq:bartheta}
\bar{\Theta}_2(q) =
  \sqrt{\frac{q}{\pi}}\,\Theta_2(q)
\end{equation}
and $\Theta_2$ is the elliptic function
\begin{equation}
  \label{eq:theta2}
\Theta_2(q) = \sum_{k=-\infty}^\infty q^{\left(k+\frac{1}{2}\right)^2}\,.
\end{equation}
Notice that $\bar{\Theta}_2(q)\to 1$ as $q\to 1$, which shows that the
$T=0$ result is recovered in the $T\to 0$ limit.
Since in the limit $B/T^2\to\infty$ the $T$-dependent part reduces to its
$T\to 0$ limit, the same asymptotic behavior is obtained for $D^S(B)$
as in the $T=0$ case. Intuitively this is clear since then $B$ is the
largest scale, which cannot be spoiled by any finite $T$. 
Using the gradient flow, the finite-temperature results are
\begin{equation}
  \label{eq:finiteT_flow}
  \begin{aligned}
    \expv{\bar\psi\psi}_B(t) 
&= \frac{m qB}{(2\pi)^2} e^{-2t(m^2+qB)}
\int_0^\infty ds\,
e^{-s m^2}
\frac{1+ e^{-2qBs}}{
1- e^{-2qB(2t+s)}}\,\frac{1}{2t+s}\,
\bar{\Theta}_2\left(e^{-(2t+s) (2\pi T)^2}\right)
\,,\\
\expv{\bar\psi\psi}_B^{LLL}(t) &=        
 \frac{ m qB}{(2\pi)^2} e^{-2t(m^2+qB)}
\int_0^\infty ds\,
e^{-s m^2}
\,\frac{1}{2t+s}\,
\bar{\Theta}_2\left(e^{-(2t+s) (2\pi T)^2}\right)\,,
  \end{aligned}
\end{equation}
which leads again to $C^S(B)\to 1$ as $B\to \infty$. 
  
Finally we consider the ratios $C^T$ and $D^T$ for the spin polarization $\expv{\psi\sigma_{xy}\psi}_B$. 
As shown in App.~\ref{app:2}, this observable is special in the free case, in the sense that it is 
made up exclusively by the LLL-contribution. Once this quantity is properly renormalized (either 
via the gradient flow or via the construction of Eq.~(\ref{eq:DTf})), the ratios become unity. 
Thus, $C^T=D^T=1$ trivially for free quarks, for any magnetic field.

\bibliographystyle{jhep_new}
\bibliography{LLL}

\end{document}